%% file: 10644_p.tex
\documentclass{aa}
\usepackage{graphicx}
\usepackage{txfonts}
\usepackage{natbib}
\usepackage[german,english]{babel}
\usepackage{psfig,lscape}
\usepackage{epsfig}
\usepackage{xspace}
\usepackage{epsf}
\newcommand{\ltapprox}{\raisebox{-0.5ex}{$\,\stackrel{<}{\scriptstyle\sim}\,$}}
\newcommand{\gtapprox}{\raisebox{-0.5ex}{$\,\stackrel{>}{\scriptstyle\sim}\,$}}
\newcommand{\zphot}{$z_{phot}$~}

\newcommand{\zspec}{$z_{spec}$~}
\newcommand\bvr{$(B-V)_{rest}$\xspace}

\newcommand\zp{$z_{phot}$\xspace}

\newcommand\zs{$z_{spec}$\xspace}

\newcommand\pz{$P(z)$\xspace}

\newcommand\pclust{$P_{clust}$\xspace}
\newcommand\pthresh{$P_{thresh}$\xspace}
\newcommand\zclust{$z_{clust}$\xspace}

\def\aap{A\&A}

\def\aj{AJ}
\def\apj{ApJ}
\def\apjl{ApJ}
\def\apjs{ApJS}
\def\mnras{MNRAS}

\begin{document} 
% ----------------------------------------------------
% ----------------------------------------------------

\title{Photometric Redshifts and cluster tomography in the ESO Distant Cluster Survey
\thanks{Based on observations collected at the European Southern Observatory,
Paranal and La Silla, Chile, as part of the ESO LP 166.A-0162. 
% Also based on observations collected with  
%NASA/ESA $Hubble$ $Space$ $Telescope$, obtained at the Space Telescope Science
%Institute, which is operated by AURA Inc, under NASA contract NAS
%5--26555.
}}
\author{R.~Pell\'o\inst{1}
\and
G.~Rudnick\inst{2} \and
G.~De~Lucia\inst{3,18} \and
L.~Simard\inst{4} \and
D.~I.~Clowe\inst{5} \and
P.~Jablonka\inst{6,7} \and 
B.~Milvang-Jensen\inst{8,19} \and
R.~P.~Saglia\inst{9}   \and
S.~D.~M.~White\inst{3} \and
%%
%% other photo-spectro builders (alphabetic)
%%
A.~Arag{\' o}n-Salamanca\inst{10} \and 
C.~Halliday\inst{11,20} \and 
B.~Poggianti\inst{12} \and
%S.~Poirier\inst{9} \and 
%%
%% other co-Is (alphabetic)
%%
P.~Best\inst{13} \and 
J.~Dalcanton\inst{14} \and 
M.~Dantel-Fort\inst{15} \and 
B.~Fort\inst{15} \and 
A.~von der Linden\inst{3}  \and
Y.~Mellier\inst{15} \and 
H.~Rottgering\inst{16} \and 
D.~Zaritsky\inst{17}
}
\offprints{R. Pell\'o, roser@ast.obs-mip.fr}
\institute{
%1
Laboratoire d'Astrophysique de Toulouse-Tarbes, CNRS, Universit\'e de
Toulouse, 14 Avenue Edouard Belin, 31400-Toulouse, France
\and
%2
Leo Goldberg Fellow,
National Optical Astronomical Observatory, 950 North Cherry Avenue, Tucson, AZ
85721, USA
\and
%3
Max-Planck-Institut 
\selectlanguage{german}f{\"u}r Astrophysik,
Karl-Schwarschild-Str. 1, Postfach 1317, D-85741 Garching, Germany
\selectlanguage{english}
\and
%4
Herzberg Institute of Astrophysics, National Research Council of
Canada, Victoria, BC V9E 2E7, Canada
\and
%5
Department of Physics and Astronomy, Ohio University, Athens, OH 45701, USA
\and
%6
Observatoire de Gen\`eve, Laboratoire d'Astrophysique Ecole Polytechnique
Federale de Lausanne (EPFL), CH-1290 Sauverny, Switzerland
\and
%7
GEPI, CNRS-UMR8111, Observatoire de Paris, section de Meudon, 5 Place
Jules Janssen, F-92195 Meudon Cedex, France
\and
%8
Dark Cosmology Centre, Niels Bohr Institute, University of Copenhagen,
Juliane Maries Vej 30, 
\selectlanguage{german}
DK--2100 Copenhagen {\O}, Denmark
%Max-Planck Institut f{\"u}r extraterrestrische Physik,
%Giessenbachstrasse, D-85748 Garching, Germany
\selectlanguage{english}
\and
%9
Max-Planck-Institut 
\selectlanguage{german}
f{\"u}r extraterrestrische Physik, 
Giessenbachstra\ss e,
Postfach 1312, 85741 Garching, Germany
\selectlanguage{english}
\and
%10
School of Physics and Astronomy, University of Nottingham,
University Park, Nottingham NG7 2RD, United Kingdom
\and
%11
Osservatorio Astrofisico di Arcetri, Largo E. Fermi 5, 50125 Firenze, Italy
\and
%12
Osservatorio Astronomico, vicolo dell'Osservatorio 5, 35122
Padova, Italy
\and
%13
SUPA, Institute for Astronomy, Royal Observatory, Blackford Hill, Edinburgh,
EH9 3HJ, UK
\and
%14
Astronomy Department, University of Washington, Box 351580, Seattle, WA 98195, USA 
\and
%15
Institut d'Astrophysique de Paris, 98 bis boulevard Arago, 75014 Paris, France 
\and
%16
Sterrewacht Leiden, P.O. Box 9513, 2300 RA, Leiden, The Netherlands
\and
%17
Steward Observatory, University of Arizona, 933 North Cherry Avenue, Tucson, AZ 85721
\and
%18
INAF - Astronomical Observatory of Trieste, via G.B. Tiepolo 11, 34143
Trieste, Italy 
\and
%19
The Royal Library / Copenhagen University Library, Research Dept.,
Box 2149, DK--1016 Copenhagen K, Denmark
\and
%20
Department of Physics and Astronomy, University of Glasgow, Glasgow G12
8QQ, United Kingdom
}
\date{\today}
\authorrunning{R. Pell\'o et al.}{ }
\titlerunning{Photometric Redshifts and cluster tomography in EDisCS}

% \abstract{}{}{}{}{}
\abstract
% 5 {} token are mandatory
% context heading (optional)
{This paper reports the results obtained on the photometric redshifts
  measurement and accuracy, and cluster tomography in the ESO Distant
  Cluster Survey (EDisCS) fields.}
% aims heading (mandatory)
{We present the methods used to determine photometric redshifts to
  discriminate between member and non-member galaxies and reduce the
  contamination by faint stars in subsequent spectroscopic studies.}
% methods heading (mandatory)
{Photometric redshifts were computed using two independent codes both
  based on standard spectral energy distribution (SED) fitting methods
  ($Hyperz$ and G. Rudnick's code).  Simulations were used to
  determine the redshift regions for which a reliable determination of
  photometric redshifts was expected. The accuracy of the photometric
  redshifts was assessed by comparing our estimates with the
  spectroscopic redshifts of $\sim 1400$ galaxies
  in the $0.3\le z\le 1.0$ domain.  
The accuracy
  expected for galaxies fainter than the spectroscopic control sample
  was estimated using a degraded version of the photometric catalog
  for the spectroscopic sample.  }
% results heading (mandatory)
{The accuracy of photometric redshifts is typically $\sigma(\Delta
  z/(1+z))\sim 0.05 \pm 0.01$, depending on the field, the filter set,
  and the spectral type of the galaxies.
  The quality of the photometric redshifts degrades by a factor of two
  in $\sigma(\Delta z/(1+z))$ between the brightest ($I\ltapprox$22)
  and the faintest ($I\sim$24-24.5) galaxies in the EDisCS sample.
  The photometric determination of cluster redshifts in the EDisCS
  fields using a simple algorithm based on \zphot is in excellent
  agreement with the spectroscopic values, such that $\delta z
  \sim$0.03-0.04 in the high-z sample and $\delta z \sim$0.05 in the
  low-z sample, i.e. the \zphot cluster redshifts are at least a
  factor $\sim(1+z)$ more accurate than the measurements of \zphot for
  individual galaxies.
  We also developed a method that uses both photometric redshift codes
  jointly to reject interlopers at magnitudes fainter than the
  spectroscopic limit. 
 When applied to the spectroscopic sample, this method
rejects $\sim 50-90\%$ of all
  spectroscopically confirmed non-members, while retaining $\gtrsim
  90\%$ of all confirmed members.}
%conclusions heading (optional)
{Photometric redshifts are found to be particularly useful for the
  identification and study of clusters of galaxies in large surveys.
They enable efficient and complete pre-selection of cluster members for
spectroscopy, allow accurate determinations of the
cluster redshifts based on photometry alone, and provide a means of
determining cluster membership, especially for bright sources.
}


\maketitle

%%%%%%%%%%%%%%%%%%%%%%%%%%%%%%%%%%%%%%%%%%%%%%%%%%%%%%%%%%%%%%%%

%%%%%%%%%%%%%%%%%%%%%%%%%%%%%%%%%%%%%%%%%%%%%%%%%%%%%%%%%%%%%%%%

\section{\label{intro} Introduction}

Photometric redshifts are becoming an important tool in cosmological
studies based on large and/or deep photometric surveys. Different
studies have been devoted to the detailed analysis of photometric
redshift accuracy in different contexts (e.g. Ilbert et al.\ 2006,
Feldmann et al.\ 2006, Mobasher et al.\ 2007,
Banerji et al.\ 2008, Margoniner \& Wittman 2008, Hildebrandt et al.\
2008, Ilbert et al.\ 2009).  
The robust evaluation of the accuracy reached by photometric
redshifts requires homogeneous deep photometric data and a large
dataset of spectroscopic redshifts for the same field. Simulations can
be used to achieve uniform coverage in parameter spaces beyond the
limits of spectroscopic surveys, in particular when missing
information about certain redshift domains and/or spectroscopic types
of galaxies.

Several papers used a recalibration between data and template models
to improve the precision of photometric redshifs (e.g. Coe et al.\
2006, Ilbert et al.\ 2006, 
Feldmann et al.\ 2006, Mobasher et al.\ 2007,
Capak et al.\ 2007, Ilbert et al.\ 2009), a method that requires
a large and representative training set of spectroscopic redshifts.
However, model templates, optimized to achieve the highest possible
accuracy in a given catalog/field, are not necesarily optimal in all
cases because systematic problems in the catalog photometry could
remain unrecognized during the template calibration process. A
more robust estimate of photometric redshifts accuracy can be
achieved for large datasets acquired in different independent fields.
This is the approach used in this paper.

The ESO Distant Cluster Survey (hereafter EDisCS) is an ESO Large
Programme designed to study the evolution of cluster galaxies over a
significant fraction of cosmic time (White et al.\ 2005). The 20
clusters included in the EDisCS sample were selected from the Las
Campanas Distant Cluster Survey (LCDCS; Gonzalez et al.\ 2001) with
redshifts ranging between $\sim$0.4 and 1. More details about the
survey and the cluster selection procedure can be found in the paper
by White et al.\ (2005), and the EDisCS
website\footnote{http://www.mpa-garching.mpg.de/galform/ediscs}.  The
EDisCS programme includes homogeneous and deep photometry with ESO VLT
and NTT (optical and near-IR; White et al.\ 2005, Arag\'on--Salamanca
et al.\ in preparation) and multi-object spectroscopy with ESO VLT
(Halliday et al.\ 2004, Milvang-Jensen et al.\ 2008), as well as other
follow-up observations with HST/ACS (Desai et al.\ 2007), narrowband
H$\alpha$ imaging (Finn et al.\ 2005) and XMM data (Johnson et al.\
2006).

This paper is also intended to be the reference for the photometric
redshifts and the cluster membership criteria
adopted by the EDisCS collaboration, and used in the
different EDisCS papers dealing with cluster membership and related
quantities (e.g. De Lucia et al.\ 2004, White et al.\ 2005, Clowe et
al.\ 2006, Poggianti et al.\ 2006, De Lucia et al.\ 2007, Desai et
al.\ 2007, Rudnick et al.\ 2009). 
Photometric redshifts are particularly useful when used with cluster/structure
finding algorithms, because they help to ensure that time-consuming 
spectroscopic observations are optimized (e.g. Li \& Yee\ 2008).

The paper is organized as follows. In Sect.~\ref{catalogs}, we
summarize the characteristics of the relevant photometric and
spectroscopic data. A technical description of the photometric
redshift methods and related procedures is provided in
Sect.~\ref{photoz}. The photometric redshift accuracy in the EDisCS
survey is addressed in Sect.~\ref{dz} using three different
approaches: 1) simulations are used to determine the redshift regions
for which a reliable determination of photometric redshifts is
expected; 2) the actual quality achieved in this survey is estimated
by direct comparison between the photometric and spectroscopic
redshifts, and 3) the accuracy expected for galaxies fainter than the
spectroscopic control sample is estimated using a degraded version of
the spectroscopic sample catalog.
Section~\ref{cluster} presents the comparison between spectroscopic
and photometric determinations of cluster redshifts, as well as the
results obtained on cluster tomography in the EDisCS fields. The
photometric cluster membership criteria adopted by the EDisCS
collaboration is introduced and discussed in Sect.~\ref{member}.
Discussion and conclusions are given in Sect.~\ref{conclusions}.
The following cosmological parameters are adopted throughout this paper: 
$\Omega_{\Lambda}=0.7$, $\Omega_{m}=0.3$, and $H_{0}=70\ km\ s^{-1}\
Mpc^{-1}$. Magnitudes are given in the Vega system. 

%%%%%%%%%%%%%%%%%%%%%%%%%%%%%%%%%%%%%%%%%%%%%%%
\section{\label{catalogs} Photometric and spectroscopic data}

We use the ground-based photometric catalogs and spectroscopic
redshifts obtained by the EDisCS collaboration for 20 clusters of
galaxies with redshifts ranging between 0.4 and 1.0 (White et al.\
2005). Although the final redshift distribution of this sample is
found to be fairly uniformly distributed within this redshift
interval, the original filter set was designed to bracket the relevant
wavelength domain at z$\sim$0.5 (low-z sample) and z$\sim$0.8 (high-z
sample). Photometric redshifts and related quantities depend strongly
on the wavelength domain covered by the photometric Spectral Energy
Distributions (hereafter SED), i.e. the filter set. Throughout
the paper, we therefore retain the original division of the clusters
into the ``low-z'' and the ``high-z'' samples.

Deep optical photometric data was acquired with FORS2 at the VLT, in
BVI and VRI bands for the low-z and the high-z cluster samples,
respectively. 
The photometric depth (5$\sigma$ in 1\arcsec radius
aperture) is typically
26.4(B), 26.2(V) and 24.8(I) in the low-z sample, and  
26.5(V), 26.0(R) and 25.2(I) in the high-z sample
(see also White et al.\ 2005, Table 1).
The field of view covered by these data is 6.5' $\times$
6.5'.  Seeing conditions were excellent during all imaging
observations, ranging typically between 0.5'' and 0.8'' (see White et
al.\ 2005 for details).  Deep near-IR images were also obtained 
for almost all clusters 
with SOFI at the NTT, in K$_s$ and JK$_s$ for the low-z and the high-z
samples, respectively (details are provided by Arag\'on-Salamanca et
al.\ 2009, in preparation). 
The photometric depth (5$\sigma$) is typically 22.8 in J and 21.5 in
K$_s$.
These data cover a field of 4.2'$\times$
6.0' at low-z, and 4.2'$\times$ 5.4' at high-z.  Photometry was
performed on seeing--matched images using SExtractor v.2.2.2 (Bertin
\& Arnouts 1996). Table~\ref{zz} summarizes the filter set used to
compute photometric redshifts for each cluster in the EDisCS sample.

%% Spectroscopic control sample
Spectroscopic data in the EDisCS fields were obtained during three
observing runs with FORS2 at VLT, using the 600RI+19 grism. The
wavelength domain covered by our observations ranged between $\sim
5300$ and 9000 \AA. More details can be found in the reference papers
by Halliday et al.\ (2004) and Milvang-Jensen et al.\ (2008). The
total number of good quality spectra acquired per field ranged between
$\sim$60 and 100 for the low-z sample, and was around $\sim$100 for
the high-z sample. There are typically 30-50 confirmed members in
every cluster. 
We use objects with either secure spectroscopic redshifts 
(hereafter type 1) or medium quality, slightly tentative redshifts (hereafter
type 2) to characterize the behavior of photometric redshifts and
cluster-membership criteria. Objects with tentative redshifts ($\sim$ 50\%
secure, type 3) represent less than 2\% of the total sample and
are mostly used for illustration pusposes.  
The total number of spectroscopic redshifts available is 637(977) in the
low-z(high-z) samples, from which the total number of secure (secure $+$
slightly tentative, i.e. type 1$+$2) redshifts in the $0.3\le z\le 1.0$ domain
are 544(564) and 854(885) respectively for the low-z and high-z samples (see also
Table~\ref{zzsum}).

%% Photometric settings/catalogs: magnitudes
An important issue when deriving photometric redshifts for a given
galaxy is to construct its SED using magnitudes and corresponding
fluxes derived for identical aperture sizes in each of the bandpass 
images. Photometric SEDs were obtained from seeing-matched images
according to the following scheme. For ``isolated'' objects
(SExtractor flag $=$0), photometric redshifts were derived from
isophotal magnitudes measured within the reference I-band isophotal
region corresponding to 1.5 $\sigma$ of the local background noise.
In the case of ``crowded'' objects (SExtractor flag $\ge$0), we used
magnitudes computed within 1'' radius apertures. This scheme enabled
us to improve the SED determination in crowded regions, while
increasing the S/N for isolated galaxies.

%%% Photometric settings/catalogs: error bars
We did not use the standard SExtractor errors because these are known
to underestimate the error for dithered data where adjacent pixels are
correlated. We determined our errors instead by means of a set of
empty aperture simulations as described in White et al.(2005). The
accurate determination of the errors is important because photometric
redshifts are sensitive to photometric errors.

%% Photometric settings/catalogs: E(B-V)
A correction for Galactic extinction was also included for each
cluster field according to the E(B-V) derived from Schlegel et
al. (1998) for the center of the cluster. The E(B-V) corrections in
the EDisCS fields typically ranged between 0.03 and 0.08 magnitudes
(see Table~\ref{zz}).

%% Photometric system
Bright unsaturated stars were used as secondary standards to check the
consistency of our photometric system for deriving photometric
redshifts and, when required, to introduce small zero-point
corrections in the photometric catalogs. In practice, we compared the
color-color diagrams of observed stars in our fields with the expected
positions derived using the Pickles library (Pickles 1998). Stars at
this stage are objects selected with SExtractor stellarity index $\ge
0.95$ that belong to the stellar sequence in the I-band
isophotal-radius versus aperture-magnitude diagrams. This procedure
was particularly useful during the first photometric runs to correct
near-IR imaging data for the effects of non-photometric
conditions. Due to the addition of high-quality observations, the
zero-point corrections improved successively during the lifetime of
the project, in addition to the quality of the photometric catalogs
and related quantities, such as photometric redshifts. The results
presented here were obtained with the final version of the
photometric catalogs, for which the zero-point corrections are
negligible, apart from two cases: Cl1138-1133 ($\Delta V=$0.10 and
$\Delta J=$-0.15), and Cl1232-1250 ($\Delta J=$-0.20).
 
%% Other papers:
All the results published by the EDisCS collaboration since 2004 
were obtained with the current and final version of the photometric catalogs
used in this paper, publically available from the EDisCS
website\footnote{http://www.mpa-garching.mpg.de/galform/ediscs}. 
The last version of EDisCS photometric redshifts was obtained in April
2006. The quality of this final version with respect to the previous ones
(since 2004) is about the same in terms of accuracy (i.e. systematic offsets,
dispersion and catastrophic failures; see criteria in Sect.~\ref{dz}). 
The main differences come from the related quantities which are provided in
addition to photometric redshifts (e.g. absolute magnitudes, photometric
classification of galaxies, ...).

%%%%%%%%%%%%%%%%%%%%%%%%%%%%%%%%%%%%%%%%%%%%%%%
\section{\label{photoz} Photometric redshifts}

Photometric redshifts (hereafter \zphot) were computed using two
different codes: a modified version of the public code {\it
  Hyperz}\footnote{http://webast.ast.obs-mip.fr/hyperz/} (Bolzonella et
al.\ 2000), and the code of Rudnick et al. (2001),
with the modifications introduced by Rudnick et al. (2003) (hereafter
GR code). The two codes use different approaches based on SED fitting
procedures, as summarized below. The reader is referred to the
reference papers for a more detailed description of the codes
themselves. Here we summarize only the main relevant settings and
modifications.

{\it Hyperz} results were initially used by the EDisCS collaboration
for three main purposes: to determine a first guess for each cluster
redshift, to help in spectroscopic pre-selection, and to reduce the
contamination by faint stars during spectroscopic observations.
Subsequently, the two codes were jointly used to establish cluster
membership in magnitude-limited samples using their respective
normalized probability distributions (see Sect.~\ref{member}).

%\subsection{Hyperz settings}

We used 14 galaxy templates with {\it Hyperz}:

\begin{itemize}

\item Eight evolutionary synthetic SEDs computed with the 2003
  version of the Bruzual \& Charlot code (Bruzual \& Charlot 1993, 2003),
  spanning a grid of ages between 0.0001 and 13.5 Gyrs, with Chabrier
  (2003) IMF and solar metallicity
  (a delta burst
  -SSP-, a constant star-forming system, and 6 $\tau$-models with
  exponentially decaying SFR). 

\item A set of 4 empirical SEDs compiled by Coleman, Wu and Weedman
  (1980) (hereafter CWW) to represent the local population of
  galaxies, with fixed age, extended to wavelengths $\lambda \le
  1400$\,\AA\ and $\lambda \ge 10000$\,\AA\ using the equivalent
  Bruzual \& Charlot spectra.

\item Two starburst galaxies (SB1 and SB2) from the Kinney et
  al. (1996) template library.

\end{itemize}

The internal reddening law is taken from Calzetti (2000), and
considered as a free parameter with $A_V$ ranging between 0 and 1.5
magnitudes (E(B-V) between 0 and $\sim$ 0.45 mags).
When an object is not detected in a given filter, 
the flux in this filter is
set to zero with an error bar corresponding to the limiting magnitude
that corresponds to a S/N ratio $\sim 1$ in this filter.
Absolute magnitudes $M_B$ for galaxies were allowed between $-24 \le
M_B \le -11$, which imposed a relatively weak prior on \zphot while
preventing obvious catastrophic identifications in the case of
degenerate solutions. P(z) were normalized within the permitted
redshift interval according to this restriction.

%\subsection{GR code settings}

The GR code is based on the non-negative linear combination of
redshifted galaxy templates, for which:

\begin{itemize}

\item The set of 4 CWW empirical templates described above.

\item The starburst galaxies SB1 and SB2 from the Kinney et al.\
  (1996).

\item A 10 Myr old, single stellar population burst obtained from the
  1999 version of the Bruzual \& Charlot (1993) code with Salpeter
  (1955) IMF and solar metallicity.

\end{itemize}

The 4 CWW and starburst templates were extended from their published
short-wavelength limits (1400 and 1232 \AA\, respectively) by a  power-law
fit to the 1400-1800 and 1240-1740 \AA\ wavelength ranges, respectively,
using Bruzual \& Charlot (2003) spectra.

There are no limitations on absolute magnitude in the GR code, and the direct flux
measurements were used for all galaxies, even when an object was not
formally detected.

{\it Hyperz} \zphot were computed in the range $0 \le z \le 6$,
whereas GR ones span the $0 \le z \le 2$ range. The upper limit had a
negligible impact on the \zphot value itself and related quantities,
except in the normalization of the \zphot probability distribution
(P(z)). When deriving the cluster membership criteria presented in
Sect.~\ref{member}, P(z) computed with the two codes were normalized
within the $0 \le z \le 2$ interval, and {\it Hyperz} \zphot were also
restricted to this interval.

%%%%%%%%%%%%%%%%%%%%%%%%%%%%%%%%%%%%%%%%%%%%%%%
\subsection{\label{stargal} Photometric discrimination between galaxies, stars
and quasars}

The method used to discriminate photometrically between galaxies,
stars, and quasars, was based on {\it Hyperz}, and closely followed
the developments presented in Hatziminaoglou et al.\ (2000). For
stars, it was based on a standard SED fitting minimization about $z
\sim 0$ using the complete library of stellar templates by Pickles
(1998). The galactic E(B-V) correction was considered as a free
parameter, ranging between 0 and the corresponding value for the field
given in Table~\ref{zz}. For quasars, we used a library of synthetic
spectra similar to Hatziminaouglou et al.\ (2000), and the same
prescriptions as for galaxies, apart from the absolute magnitude
limitation.

In practice, the usual {\it Hyperz} \zphot for galaxies and quasars,
and the best fit with the stellar library were computed for each
object, and three classification parameters were given to quantify the
goodness of the best fit as a galaxy/star/quasar (respectively N$_G$,
N$_*$ and N$_Q$). The object was ``rejected'' as a galaxy/star/quasar
when its $\chi^2$ excluded it at higher than the 95\% confidence level
(N$=$0). The object was ``fully compatible'' when the probability
associated with the reduced $\chi^2$ exceeded 90\% (N$=$2). The object
was ``undetermined'' (N$=$ 1) in all the other intermediate cases.
This classification allowed us to define different samples of objects
in these fields, either galaxies (with N$_G$ $\ge 1$, irrespective of
the star type) or stars (with N$_*$ $>$ 1 and N$_G$$<$1). These
classification criteria were used during the spectroscopic runs to
lower the contamination due to faint and red stars to values below
$\sim 10\%$ (Halliday et al.\ 2004; Milvang-Jensen et al.\ 2008), and
they are also used below in \S~\ref{tomography}.  The quasar
classification was not considered for the spectroscopic preselection.
Johnson et al.\ (2006) used this classification to identify possible
AGNs detected in X-rays in these fields.

%%%%%%%%%%%%%%%%%%%%%%%%%%%%%%%%%%%%%%%%%%%%%%%
\subsection{\label{types} Photometric classification of galaxies}

Galaxies were classified into five different spectral types,
corresponding to their rest-frame photometric SED: (1) E/S0, (2) Sbc,
(3) Scd, (4) Im, and (5) SB (starbursts). These types correspond to
the simplest empirical templates given above, namely the four SEDs
compiled by CWW for the local galaxies, plus the Kinney
starbursts. 
This classification corresponds to the best fit templates for the GR
code. In the case of {\it Hyperz}, it is the best fit of the rest-frame SED at
\zphot. The classification obtained with the two codes is in perfect agreement,
excepted for catastrophic identifications.  
We use this classification in \S~\ref{dz} to address the
\zphot accuracy as a function of the spectral type.

%%%%%%%%%%%%%%%%% Figs zz simus 
\begin{figure}
{\centering \leavevmode
\psfig{file=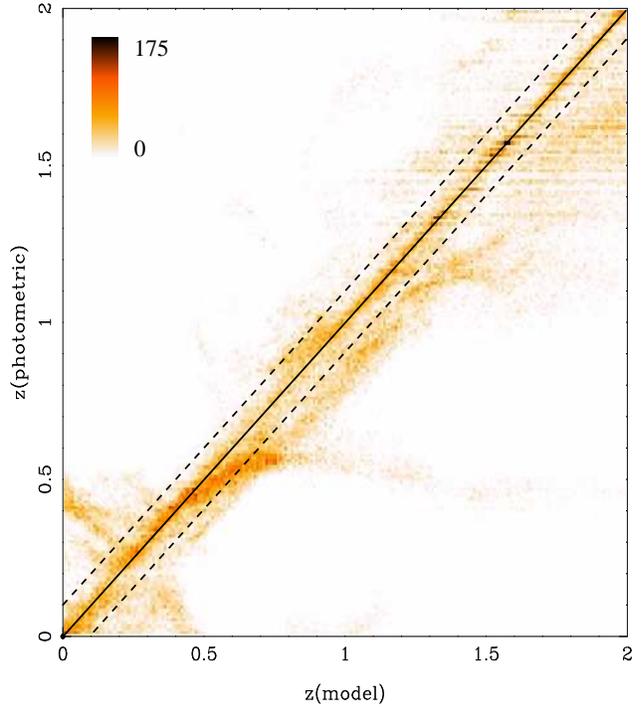,width=0.45\textwidth}
}
\caption{Photometric versus model redshifts retrieved from BVIK$_s$
  SEDs (low-z sample), for $10^5$ simulated galaxies uniformly
  distributed between $0 \le z \le 2$. The diagram displays the number
  density of galaxies in linear scale.}
\label{model_lowz}
\end{figure}

\begin{figure}
{\centering \leavevmode
\psfig{file=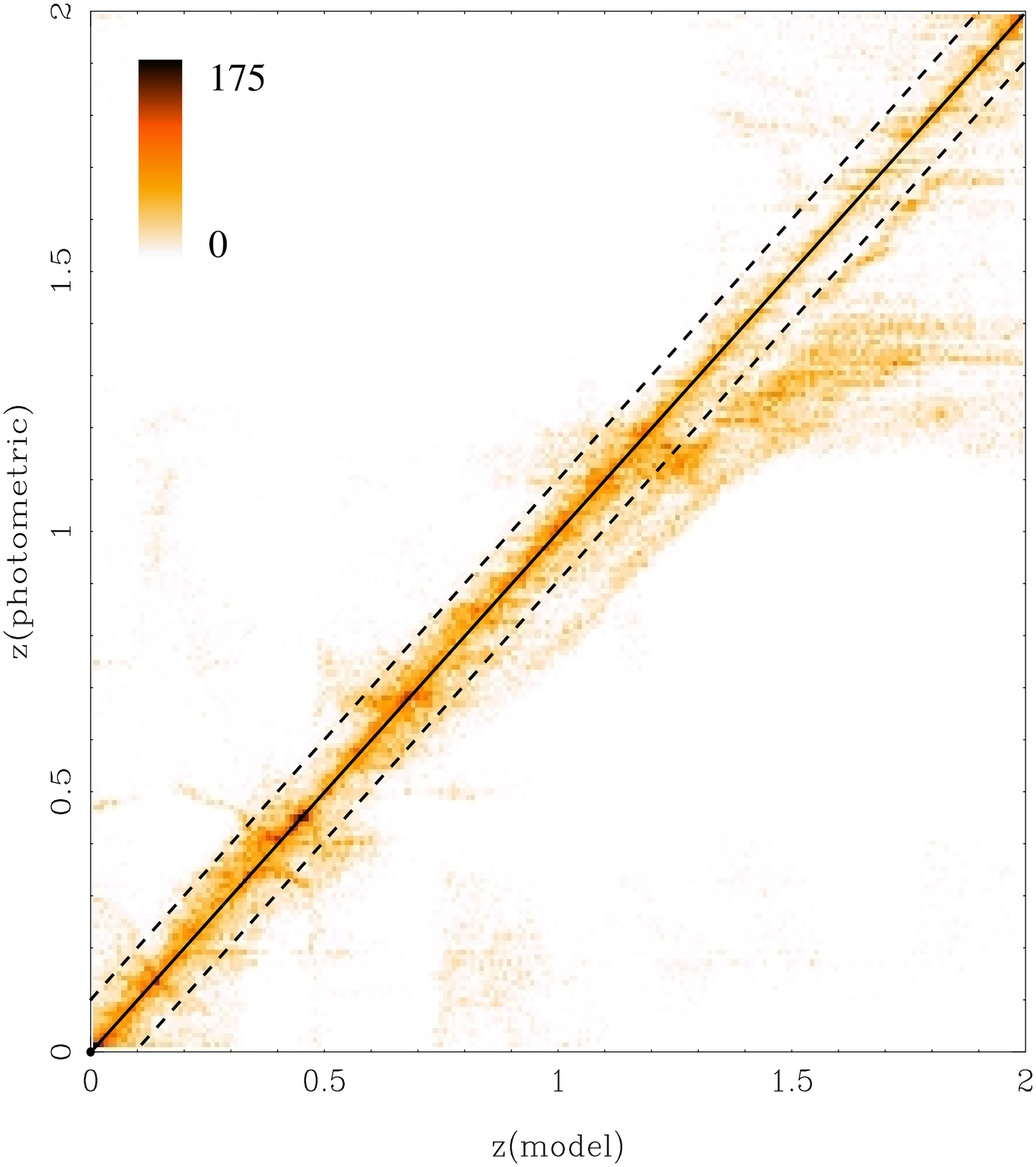,width=0.45\textwidth}
}
\caption{Photometric versus model redshifts retrieved from VRIJK$_s$
  SEDs (high-z sample), for $10^5$ simulated galaxies uniformly
  distributed between $0 \le z \le 2$. The diagram displays the number
  density of galaxies in linear scale.  }
\label{model_highz}
\end{figure}

\begin{figure}[h]
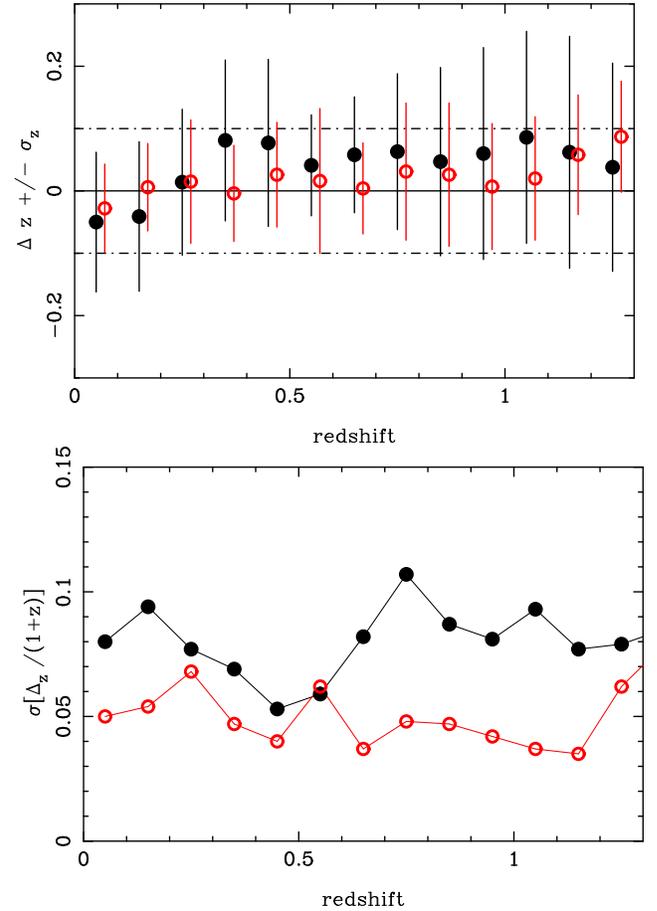

{\centering \leavevmode
\psfig{file=10644f3.ps,angle=270,width=0.45\textwidth}
\newline
\psfig{file=10644f4.ps,angle=270,width=0.45\textwidth}
\caption {These figures present the expected \zphot accuracy derived
  from simulations, for galaxies with photometric quality similar to
  the spectroscopic sample.  The top panel displays the systematic
  deviation $\left<\Delta_z \right> \pm \sigma_z$, the standard
  deviation excluding catastrophic identifications, for the BVIK$_s$
  (low-z, black dots) and VRIJK$_s$ (high-z, open/red dots) filter
  sets, within the relevant redshift domain.  The botton panel
  displays a plot of $\sigma(\Delta z/(1+z))$ as a function of
  redshift for the same filter combinations.}
\label{sigma_models}
}
\end{figure}

%%%%%%%%%%%%%%%%%%%%%%%%%%%%%%%%%%%%%%%%%%%%%%%
\subsection{\label{public} Photometric redshift catalogs}

Photometric redshifts for EDisCS catalogs computed with the two
different codes are publically available from the EDisCS
website\footnote{http://www.mpa-garching.mpg.de/galform/ediscs}.

EDisCS online catalogs also include an optimized flag for the
discrimination between galaxies and stars, based on the combination
between the above {\it Hyperz} criterion, a similar fit using the GR
code (flag$_{GR}$=1 when the object is well fit as a star) (1), the
SExtractor stellarity flag (flag$_{SEx}$) (2), and the size for bright
objects (3), i.e.

\begin{enumerate}

\item $\{$ N$_* > $1 AND  N$_G < $1 $\}$ OR flag$_{GR}=$ 1

\item flag$_{SEx} >$ 0.95

\item r$_h$(I) $<$ r$_{thresh}$ if I$_{tot} < $22.5

\end{enumerate}

where r$_h$(I) is the half-light radius in the I band, and
r$_{thresh}$ is the threshold radius determined from the stellar locus
in the corresponding r$_h$ versus total I magnitude.

%%%%%%%%%%%%%%%%%%%%%%%%%%%%%%%%%%%%%%%%%%%%%%%
\section{\label{dz} Determination of the photometric redshift accuracy}
%%%%%%%%%%%%%%%%%%%%%%%%%%%%%%%%%%%%%%%%%%%%%%%

The expected accuracy of \zphot as a function of redshift depends
strongly on the photometric accuracy and the filter set used to derive
the photometric SEDs. As explained in \S~\ref{catalogs} and in White
et al.\ (2005), we introduced an accurate determination of photometric
errors to address the former issue. The filter set used by EDisCS
contains a relatively small number of filters because it was designed
originally to cover the relevant wavelength domain in the rest frame
of low z$\sim$0.5 and high z$\sim$0.8 clusters. In particular, they
bracketed the 4000\,\AA\ break for the relevant redshift intervals,
but they were not designed to explore the full redshift domain. For
this reason, we address the \zphot accuracy in three different ways
described below. Using simulations, we first determine the redshift
ranges for which we expect a reliable estimate of \zphot for the low
and high-z filter sets, and the ideal (maximum) accuracy expected from
simple SED fitting. Secondly, the quality of EDisCS \zphot achieved is
estimated by direct comparison with the spectroscopic redshifts. In a
third subsection, we determine the \zphot accuracy expected for
galaxies fainter than the spectroscopic control sample.

The following quantities were computed to quantify the \zphot
accuracy, where \zspec stands for both ``spectroscopic'' and ``model''
redshifts:

\begin{itemize}

\item The systematic deviation between \zphot and \zspec,
  $\left<\Delta_z \right>=\sum \Delta_z /N$, given by the mean
  difference between these two quantities, where $\Delta_z$ = \zspec -
  \zphot and $N$ is the total number of galaxies.

\item The standard deviation $\sigma_z$ $ = \sqrt{\sum
    (\Delta_z-\left<\Delta_z \right>)^2/(N-1)}$, excluding
  catastrophic identifications, defined here in a conservative way for
  those galaxies with $ | \Delta_z | $= $|$\zspec - \zphot $| \ge 0.3
  \times $(1+ \zspec).

\item The median absolute deviation $\sigma_{z,MAD}$ $ = 1.48 \times
  median \\|$ \zspec - \zphot $|$, which is less sensitive to outliers.

\item The normalized median absolute deviation defined as
  $\sigma(\Delta z/(1+z))$ $ = 1.48 \times median (|$\zspec - \zphot $
  |/(1+ $\zspec)),

\item The percentage of catastrophic identifications ($l\%$), i.e.
  galaxies ``lost'' from their original spectroscopic redshift bin,
  with $| \Delta_z | \ge 0.3 \times (1+$ \zspec).

\item The percentage of galaxies included in a given photometric
  redshift interval that are catastrophic identifications ($g\%$),
  i.e. galaxies that contaminate the sample because they are
  incorrectly assigned to the redshift bin.

\end{itemize}

%%%%%%%%%%%%%%%% Table simus %%%%%%%%%%%%%%%% 
\begin{table*}
\begin{center}
\caption{\label{zzsimulated} \zphot accuracy from simulations 
for galaxies in the spectroscopic sample, based on {\it Hyperz}.}
\begin{tabular}{clccccccc}
\hline \hline
Clusters & Redshift interval & Galaxy type &
$\left<\Delta_z \right>$ & $\sigma_z$ &
$\sigma_{z,MAD}$ & $\sigma(\Delta z/(1+z))$ & $l\%$ & $g\%$  \\
\hline 
Low-z & $0.3 \le$  \zphot$ \le 1$ &  all & 
0.061   &  0.131 & 0.124  &   0.076 & 2.2 &    6.0 \\
High-z & $0.3 \le$  \zphot$ \le 1$ & all &
0.015 &    0.101 & 0.074  &   0.045 & 6.6  &   2.8 \\
High-z & $0.45 \le$  \zphot$ \le 1$ & all &  
0.019  &   0.108 & 0.080 &    0.046 & 8.2 &    2.8\\
\hline
Low-z & $0.3 \le$  \zphot$ \le 1$ &  E/S0 & 
-0.036  &   0.121 & 0.087  &   0.052 & 3.4 &    2.2 \\
      &                             & Sbc & 
0.075  &    0.108  &0.176   &   0.105 & 2.2 &     5.0 \\
      &                             & Scd & 
0.110  &    0.082 & 0.176  &    0.107 &  0.8  &    1.9 \\
      &                             & Im & 
0.023 &    0.089 & 0.086   &  0.053 & 0.4 &    5.6 \\
      &                             & SB &
0.095  &   0.124 & 0.126  &   0.075 & 0.6 &   11.6 \\
\hline
High-z & $0.3 \le$  \zphot$ \le 1$ & E/S0 & 
-0.019  &   0.040 & 0.031  &   0.019 & 0.0 &    0.2 \\
      &                             & Sbc & 
0.037  &    0.081  & 0.077  &    0.048 & 0.1   &   0.7 \\
      &                             &  Scd & 
0.030  &   0.072 & 0.070  &   0.043 & 0.0  &   1.4  \\
      &                             & Im & 
0.050   &  0.077 & 0.085  &   0.050 &0.2  &   1.1 \\
      &                             & SB & 
-0.022  &   0.127 & 0.087  &   0.054 &7.4  &   3.5 \\
\hline 
\end{tabular}
\end{center}

Notes -- The accuracy reached for
  the different spectral types of galaxies is also presented. The
  information given is the same as in Table~\ref{zzsum}, for the
  current spectroscopic sample.  
\end{table*}

%%%%%%%%%%%%%%%% end Table simus %%%%%%%%%%%%%%%% 

%%%%%%%%%%%%%%%% Table faint %%%%%%%%%%%%%%%% 
%% after correction for errors in input catalog

\begin{table*}
\begin{center}
\caption{\label{tab_zzfaint} \zphot accuracy expected for the faintest galaxies in the
  EDisCS sample, in the low-z and high-z fields, based on {\it Hyperz}. }
\begin{tabular}{clccccccc}
\hline \hline
Clusters & Redshift interval & Galaxy type &
$\left<\Delta_z \right>$ & $\sigma_z$ &
$\sigma_{z,MAD}$ & $\sigma(\Delta z/(1+z))$ & $l\%$ & $g\%$  \\
\hline 
Low-z & $0.3 \le$  \zphot$ \le 1$ &  all & 
-0.003  &   0.153& 0.158 &    0.105 &4.0 &    1.0 \\
High-z & $0.3 \le$  \zphot$ \le 1$ & all &
-0.037  &    0.159&  0.156  &    0.095 &  2.2 &   6.2 \\
High-z & $0.45 \le$  \zphot$ \le 1$ & all &  
-0.027 &     0.156 & 0.151 &     0.091 & 1.0 & 6.9 \\
\hline
Low-z & $0.3 \le$  \zphot$ \le 1$ &  E/S0 & 
-0.025 & 0.155 & 0.172 & 0.113 & 3.2 & 0.0  \\
      &                             & Sbc & 
0.025 & 0.151 & 0.158 & 0.106 & 4.6 & 0.7 \\
      &                             & Scd & 
0.021& 0.129 & 0.124 & 0.082 & 2.7 & 1.8 \\
      &                             & Im & 
-0.012 & 0.172 & 0.161 & 0.113 & 8.8 & 2.5  \\
      &                             & SB &
-0.096& 0.156 & 0.238 & 0.155 & 6.7 & 5.6 \\
\hline
High-z & $0.3 \le$  \zphot$ \le 1$ & E/S0 & 
-0.040  &   0.155 &0.151&     0.091 &2.0& 2.6 \\
      &                             & Sbc & 
-0.041 &     0.164&  0.172 &     0.108 & 2.1 & 1.9  \\
      &                             &  Scd & 
-0.017 &    0.166 &0.146 &    0.092 & 1.7 & 8.7 \\
      &                             & Im & 
-0.026  &   0.143& 0.142 &    0.088 & 0.9 & 2.8 \\
      &                             & SB & 
-0.075 &    0.150& 0.191  &   0.125 & 6.3 & 23.7 \\
\hline 
\end{tabular}
\end{center}

Notes -- 
The accuracy reached for the different spectral types of
  galaxies is also presented.  The information given is the same as in
  Table~\ref{zzsum}, for the current spectroscopic sample.
\end{table*}

%%%%%%%%%%%%%%%% end Table faint %%%%%%%%%%%%%%%% 
%%%%%%%%%%%%%%%%%%%%%  Figures zz %%%%%%%%%%%%%%%%%%%%%%%%%%%%%%%%%
\begin{figure*}[!ht]
{\centering \leavevmode
\psfig{file=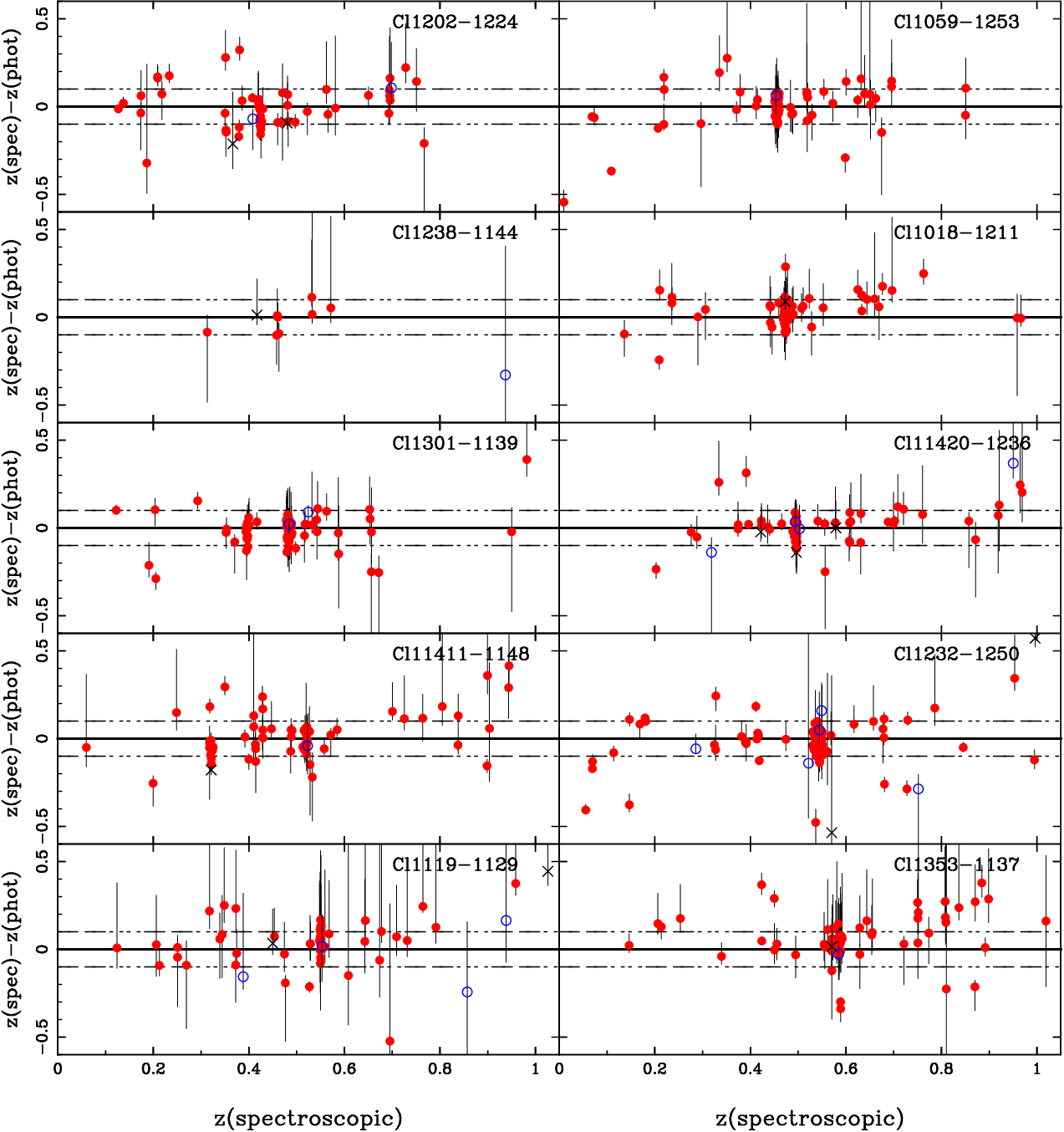,width=1.0\textwidth}
}
\caption{
(\zspec -  \zphot) versus spectroscopic
  redshift for the low-z clusters in the EDisCS sample. Solid (red)
  circles, open (blue) circles, and crosses correspond to objects with
  good (type 1), medium (type 2) and tentative (type 3) spectroscopic
  redshift determinations, respectively. Error bars in \zphot
  correspond to 1$\sigma$.
Dot-dashed lines display \zspec -  \zphot $= \pm 0.1$ to guide the eye.
(see text for details). }
\label{zz_lowz1}
\end{figure*}

\begin{figure*}[!ht]
{\centering \leavevmode
\psfig{file=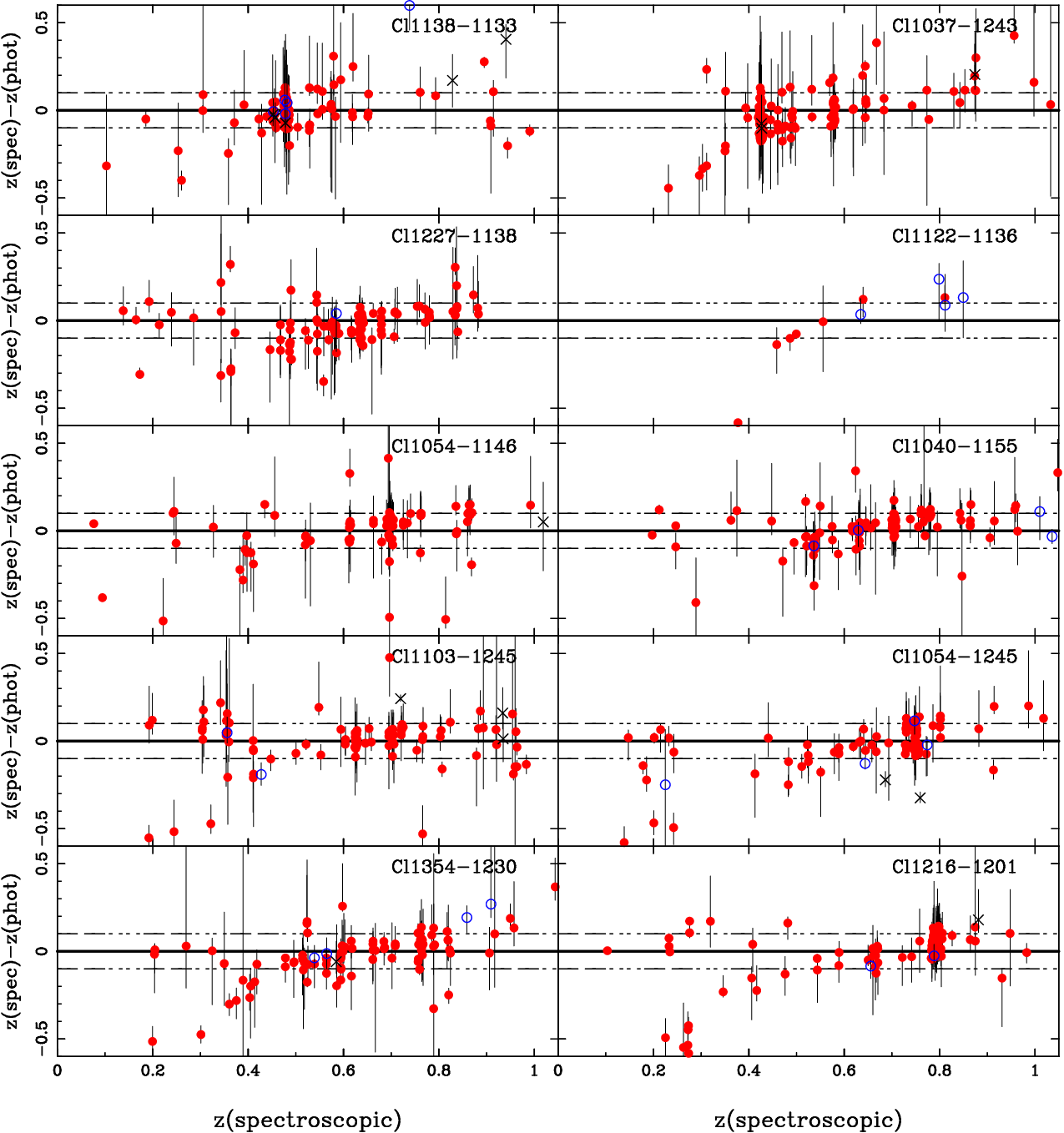,angle=270,width=1.0\textwidth}
}
\caption{ (\zspec -  \zphot) versus spectroscopic
  redshift for the high-z clusters in the EDisCS sample. Solid (red)
  circles, open (blue) circles, and crosses correspond to objects with
  good (type 1), medium (type 2) and tentative (type 3) spectroscopic
  redshift determinations, respectively. Error bars in \zphot
  correspond to 1$\sigma$ 
Dot-dashed lines display \zspec -  \zphot $= \pm 0.1$ to guide the eye.
(see text for details).}
\label{zz_highz1}
\end{figure*}

%%%%%%%%%%%%%%%%%%%%%  END Figures zz %%%%%%%%%%%%%%%%%%%%%%%%%%%%%%%%%

%%%%%%%%%%%%%%%%%%%%%%%%
\subsection{\label{simus} Expected accuracy from simulations}

Photometric redshift determinations are based on the detection of
strong spectral features, such as the $4000$\,\AA\ break, Lyman break
or strong emission lines. In general, broad-band filters allow only
detection of strong breaks and are insensitive to the presence of
emission lines, apart from when the contribution of a line to the
total flux in a given filter is higher than or similar to the
photometric errors, as happens in the case of AGNs (Hatziminaoglou et
al. 2000).

To determine the redshift domains where a reliable measurement of
\zphot can be obtained given the filter sets used in the low and
high-z samples, we completed a series of simulations assuming a
homogeneous redshift distribution in the redshift interval $0 \le z
\le 2$. These simulations were performed using 
{\it Hyperz} related software, and \zphot computed with both {\it Hyperz}
and the GR code, 
but the results should be representative of the
general behavior of pure SED-fitting \zphot codes. Synthetic catalogs
contain $10^5$ galaxies within this redshift interval, for each filter
set, spanning all the basic spectral types defined in \S~\ref{types},
with uniform redshift distribution. Photometric errors in the
different filters were assigned following a Gaussian distribution with
$\sigma$ scaled to magnitudes according to $\Delta m \simeq 2.5 \log
[1+1/(S/N)]$, where $S/N$ is the signal-to-noise ratio, which is a
function of apparent magnitude. Here we used the mean $S/N$ achieved
in the different filters for the spectroscopic (bright) sample,
i.e. for objects ranging between I$=$18.5(19.0) and 22.0 in the low-z
(high-z) sample, and the same settings used for \zphot
computation on real catalogs.  In this way, the results obtained from
simulations should be considered to be ``ideal'' but still consistent
with those derived in \S~\ref{zzspectro}. Because of the limited scope
of these simulations, we do not intend to explore all possible domains
of parameter space, but focus instead on studying the main systematics
introduced by the photometric system.

Figures ~\ref{model_lowz} and ~\ref{model_highz} display the
photometric versus model redshifts retrieved from BVIK$_s$ (low-z
sample) and VRIJK$_s$ (high-z sample), respectively. Figure
~\ref{sigma_models} and Table~\ref{zzsimulated} summarize the quality
of \zphot in the simulations using {\it Hyperz}, 
within the relevant redshift domain.
The same results are presented in Table~\ref{GR_zzsimulated} for the GR
code.

Some systematic trends are clearly visible in figures ~\ref{model_lowz},
~\ref{model_highz} and ~\ref{sigma_models}. 
The high-z filter combination provides a higher quality and smaller
systematic deviations than the low-z one. This trend was expected
because of the more complete and contiguous spectral coverage of the
high-z set. The lack of an R filter for the low-z sample introduces a
systematic trend in $\Delta_z$ $\sim$ 0.05-0.08 at $z \gtapprox 0.3$;
the highest \zphot quality for this sample is expected to be around
$\sigma(\Delta z/(1+z)) \ltapprox 0.06$ at $0.4 \ltapprox z \ltapprox
0.6$, i.e. within the sensitive redshift domain. Because of the lack
of B-band photometry for the high-z sample, the highest quality
results are expected at $z \gtapprox 0.4$. This trend is indeed
observed in the simulations. 
However, the quality achived for high-z
simulated data at $z \ltapprox 0.4$ with {\it Hyperz}
is expected to be overestimated
with respect to real data, because templates and models are drawn from
the same parent set. This is a general criticism of simulations used
in assessing the realistic performance of \zphot quality. Also \zphot
quality depends on the spectral type of galaxies. With respect to the
average quality presented in Fig. ~\ref{sigma_models} for a uniform
distribution of types, early types exhibit up to a $\sim$50\%
improvement in $\sigma(\Delta z/(1+z))$ with {\it Hyperz}
in the redshift domains where the filter sets bracket the 4000\,\AA\ break.  

The results for the GR code in Table~\ref{GR_zzsimulated} are similar in
average to {\it Hyperz}'s ones. The quality tend to be slightly better for the
bluest spectral types, whereas it is worse for early types. This trend can
be explained by the broad paramer space spanned by the simulations, the same
used by {\it Hyperz} for \zphot determinations, as compared to the GR
code. The noise for late-type galaxies in {\it Hyperz} tend to be dominated by
degeneracies, whereas the GR code cannot properly fit highly-reddened
E-Sbc galaxies. Note that the uniform distribution in redshift and types in
these simulations does not represent a realistic population of galaxies.

%%%%%%%%%%%%%%%%%%%%%%%%%%%%%%%%%%%%%%%%%%%%%%%%%%%%%%%%%%%%%%%%%%%%%%%
%%%%%%%%%%%%%%%%%%%%%%%%
\subsection{\label{zzspectro} Comparison with spectroscopic redshifts}

The \zphot accuracy achieved for EDisCS was estimated by direct
comparison with its 
1449 spectroscopic redshifts in the $0.3 \le z \le 1$ interval
(type 1$+$2).
Although
spectroscopic targets in our sample were strongly biased in favor of
cluster members with \zphot within the interval z$_{cluster} \pm 0.2$
(see Halliday et al.\ 2004; Milvang-Jensen et al.\ 2008, and
\S~\ref{preselection}), the geometrical configuration of slits in one
hand, and the need for reference field galaxies on the other hand,
ensured that non-member galaxies were also targeted during the
spectroscopic runs. In principle, these field galaxies should allow us
to extend the present study to the $0 \le$ \zspec $\le 1$ interval,
according to the restrictions imposed by the set of filters (see
\S~\ref{simus}).

Table~\ref{zz} presents the \zphot accuracy obtained with {\it Hyperz}
and G. Rudnick's (GR) code for the different low-z and high-z
clusters, based on the direct comparison with the spectroscopic
sample, excluding stars.  Only type 1 objects, i.e. objects with
secure spectroscopic redshifts, were taken into account. In
Table~\ref{zz}, systematic deviations and $\sigma_z$ were computed
over all the $0.3 \le z_{spec} \le 1$ interval.  In the low-$z$ bin,
the two clusters with only BVI photometry were excluded from the
sample when computing the average \zphot quality over the cluster
sample. In the high-$z$ bin, we excluded Cl1122-1136 from the cluster
statistics, because of the small number of \zspec available in this
field.

Our main result is that there is no significant {\it systematic}
shift, neither in the $\left<\Delta_z \right> \pm \sigma_z$ nor in the
$\sigma(\Delta z/(1+z))$ results, with respect to the values expected
from simulations in \S~\ref{simus}, with some field-to-field
differences discussed below.  The accuracy of \zphot ranges usually
between $\sigma(\Delta z/(1+z))\sim0.05$ for {\it Hyperz} and
$\sim0.06$ for GR, both for the low-z and the high-z samples. This
result is in good agreement with the highest possible accuracy
expected from ideal simulations in the low-z case, 
it is $\sim$ 25\% worse than ideal expectations in the high-z case
for {\it Hyperz}, and compatible with GR results for late type galaxies 
(early-type errors were overestimated, as mentioned in Sect.~\ref{simus}).
The differential trend between low and high-z samples with respect to
simulations in both codes 
can be explained because of the different population of
``bright'' galaxies in these fields, low-z and high-z samples
containing a smaller and larger fraction of late-type galaxies
respectively, compared to the uniform average population in simulated
data (e.g. De Lucia et al.\ 2007). This effect is clearly seen in
Table~\ref{zzsum}.
The fraction of objects lost from (or spuriously assigned to) the
relevant redshift interval according to the definitions given above
($l\%$ and $g\%$), is negligible in the low-z sample and typically
below 5\% for the high-z one.

Table~\ref{zzsum} summarizes the results obtained for the entire low-z
and high-z samples, for both type 1 (secure) and type 1 $+$ type 2
(both secure and slightly tentative redshifts) spectroscopic data.
The results are similar in both cases. In the high-z bin, \zphot
accuracy improves slightly when the sample is restricted to the $0.45
\le z_{spec} \le 1$ interval, where both {\it Hyperz} and GR codes
yield the same $\sigma(\Delta z/(1+z))\sim0.052$. This effect is
easily understood because at z$\ltapprox$0.45 the rest-frame
4000\,\AA\ break is found shortward of the V-band filter.
Table~\ref{zzsum} also summarizes the \zphot accuracy achieved for the
different spectral types of galaxies in the entire sample,
i.e. all type 1, 2 and 3 spectra. Tentative type 3 galaxies 
represent less than 2\% of the total sample, and the results remain unchanged
whith respect to type 1$+$2. The number of galaxies as a function of the
spectral type given in this table correspond to {\it Hyperz}. Excepted for
catastrophic identifications, the classification obtained with the two codes
is in perfect agreement. 
The lowest
quality $\sigma(\Delta z/(1+z))$ values are measured as expected for
the bluest galaxy types (SB), for both the {\it Hyperz} and GR
codes. Early types display the highest quality results with {\it
  Hyperz}, whereas GR code has lower quality results for early types
in the high-z sample. This trend may indicate that the CWW templates
provide an inappropriate description of the SEDs of early types at
intermediate redshifts.

The comparison between the {\it Hyperz} and GR codes, either on a
cluster-by-cluster basis or as a function of the filter combination,
yields similar results, even though these codes are based on different
approaches and have different strengths/weaknesses. In general, {\it
  Hyperz} results are found to be of slightly higher accuracy than
GR's ones (by $\ltapprox$20\% in $\sigma(\Delta z/(1+z))$), but both
are in close agreement with the expectations under ``ideal''
conditions. An interesting trend is that the quality of both codes is
highly correlated, in the sense that the highest and the lowest
quality results (in terms of $\sigma(\Delta z/(1+z))$ and systematics)
are found for the same clusters.  Given the homogeneous photometry of
the EDisCS project, this trend can hardly be explained by the use of
an incomplete or imperfect template set (as suggested by
Ilbert et al.\ 2006), because in such a case we would expect the same
systematic behavior in all fields, given a certain filter set, as is
observed in \S~\ref{simus} with {\it Hyperz}. In contrast, different
systematic trends are observed for different clusters, which are then
found to be almost identical for the two independent \zphot
codes. This behavior suggests strongly that the origin of the
systematics is more likely to be the input photometric catalog rather
than the \zphot codes and templates. In particular, we cannot exclude
small remaining zero-point shifts in our data, approximately equal to
or less than $\sim$ 0.05 magnitudes, because we are limited by the
accuracy of the stellar templates (see Sect.~\ref{catalogs}).

A brief discussion of particular aspects of \zphot accuracy in the
low-z and high-z samples is given below.

%%%%%%%%%%%%%%%%%%%%%%%%%%%%%%%%%%%%%%%%%%%%%%%
\subsubsection{Low-z cluster fields}

Figure~\ref{zz_lowz1} displays a direct comparison between the
spectroscopic and the photometric redshifts for the low-z clusters in
the EDisCS sample. {\it Hyperz} was used to derive \zphot in this
figure, but the results with the GR code are very similar, as
discussed above. Error bars in \zphot correspond to a 1$\sigma$
confidence level in the photometric redshift probability distribution
\pz, i.e. to the 68\% confidence level computed through the $\Delta
\chi^2$ increment for a single parameter (Avni 1976).
Figure~\ref{zz_low} shows the comparison between spectroscopic and
photometric redshifts for the entire low-$z$ sample, obtained with
{\it Hyperz} and GR codes, as well as the $z_{\rm spec} - z_{\rm
  phot}$ distribution.

The \zphot quality in this sample ranges usually between $0.04
\ltapprox \sigma(\Delta z/(1+z)) \ltapprox 0.07$ with both {\it
  Hyperz} and GR codes, with some exceptions. On the one hand,
Cl1119-1129 and Cl1238-1144 were only observed in BVI, which produces
lower quality \zphot and a higher fraction of catastrophic
identifications. 
These two clusters were not included when deriving
the mean values in Table~\ref{zz}. 
On the other hand, Cl1232-1250
was observed in J in addition to BVIK, and this provides a more
accurate \zphot estimate with respect to average
with {\it Hyperz}, although there is no clear improvement with the GR
code. 
Compared to simulations, the systematic trend $\Delta_z$ $\sim$
0.05-0.08 at $z \gtapprox 0.3$ is far smaller in real data, whereas
$\sigma(\Delta z/(1+z))$ is in agreement with ideal results.

%%%%%%%%%%%%%%%%%%%%%%%%%%%%%%%%%%%%%%%%%%%%%%%
\subsubsection{High-z cluster fields}

Figure~\ref{zz_highz1} displays a direct comparison between the
spectroscopic and the photometric redshifts for the high-z clusters in
the EDisCS sample. Error bars in \zphot correspond to 1$\sigma$
confidence level in the photometric redshift probability distribution
\pz.
Figure~\ref{zz_high} shows the comparison between spectroscopic and
photometric redshifts for the whole high-$z$ sample, obtained with
{\it Hyperz} and GR codes, as well as the $z_{\rm spec} - z_{\rm
  phot}$ distribution.

The \zphot quality in this sample usually ranges between $0.05
\ltapprox \sigma(\Delta z/(1+z)) \ltapprox 0.08$ with both {\it
  Hyperz} and GR codes, with some exceptions. The statistics in
Cl1122-1136 is based on a small number of spectroscopic redshifts,
hence we exclude this cluster when deriving the mean values in
Table~\ref{zzsum}.
Two out of the ten clusters in the high-z sample are actually 
in a redshift range typical of the low-z sample. Indeed, 
in the case of 
Cl1037-1243a
and Cl1138-1133, the low redshift of the
cluster implies that B-band photometry is required to ensure that an
accurate \zphot measurement is achieved, although the quality of their
\zphot measurements is close to average. As seen in
Fig.~\ref{zz_highz1}, individual error bars are larger in these two
fields than in the other high-z clusters. Compared to simulations,
there is no systematic trend in $\Delta_z$ as expected, whereas
$\sigma(\Delta z/(1+z))$ is in agreement with ideal results.

%%%%%%%%%%%%%%%%%%%%%%%%%%%%%%%%%%%%%%%%%%%%%%%%%%%%%%%%%%%%%%%%%%%%%%
%%%%%%%%%%%%%%%%%%%%%%TABLES LANDSCAPE%%%%%%%%%%%%%%%%%%%%%%%%%%%%%%%%
\clearpage
\onecolumn
\landscape
\centering
\setlength\topmargin{0cm}
\setlength\textheight{11cm}
\begin{table}[p]
\caption{\label{zz} Summary of \zphot accuracy achieved for the
individual low-z and high-z clusters.}
\include {10644_t1}

Notes -- The information given in this table
is: (1) cluster identification, (2) filter set used to compute \zphot, 
(3) spectroscopic z$_{cluster}$, (4) E(B-V) at
the cluster center, (5) total number of redshifts available in the field, 
(6) total number of spectroscopic redshifts used for 
accuracy determinations, and \zphot accuracy achieved with {\it Hyperz}
and GR codes: (7)(13) systematic deviation between \zphot and \zspec,
(8)(14) standard deviation $\sigma_z$, 
(9)(15) median absolute deviation ($\sigma_{z,MAD}$),
(10)(16) normalized median absolute deviation ($\sigma(\Delta z/(1+z))$),
(11)(17) fraction of catastrophic identifications ($l\%$),
(12)(18) fraction of spurious identifications ($g\%$). Clusters excluded in
the computation of the averaged accuracy are given in italic (see text). 
The spectroscopic z$_{cluster}$ in column (3) corresponds to the most
prominent cluster in the field. Other clusters have been identified in 
cl1037 ($z=$0.578),
cl1103 ($z=$0.959 and 0.626),
cl1138 ($z=$0.455),
cl1227 ($z=$0.583),
cl1301 ($z=$0.397), and
cl1354 ($z=$0.595) (see Milvang-Jensen et al.\ 2008 for details).
\end{table}

\setlength\topmargin{0cm}
\setlength\textheight{10cm}
\begin{table}[p]
\caption{\label{zzsum} Summary of \zphot accuracy achieved for the
whole low-z and high-z samples with {\it Hyperz} and GR codes.}
\include {10644_t2}

Notes -- 
Results for type 1 and type 1$+$2
spectroscopic data are presented in this table, as well as the accuracy
reached for the different spectral types of galaxies in the whole sample 
(i.e. all type 1, 2 and 3 redshifts). 
The information given is the same as in Table ~\ref{zz}.
\end{table}

\endlandscape
\twocolumn
%%%%%%%%%%%%%%%%%%%%%%END TABLES LANDSCAPE%%%%%%%%%%%%%%%%%%%%%%%%%%%%%%%%
%%%%%%%%%%%%%%%%%%%%%%%%%%%%%%%%%%%%%%%%%%%%%%%
\subsection{\label{zzfaint} Expected accuracy for galaxies fainter than the
  spectroscopic sample}

We determine the \zphot accuracy expected for galaxies fainter than
the spectroscopic control sample used in \S~\ref{zzspectro},
i.e. galaxies with magnitudes typically ranging between I$=$18.5(19.0)
and 22.0 in the low-z (high-z) sample, in particular for the faintest
galaxies in the EDisCS sample. The concept is to derive \zphot on a
degraded version of the photometric catalog for the spectroscopic
sample (in terms of $S/N$), using the same recipes and settings as the
main catalogs. This method was preferred instead of simulations
because it uses the observed SEDs of the control sample instead of an
arbitrary mixture of spectral types at a given redshift.

Degradated catalogs were generated from the original (spectroscopic)
ones, to reproduce the photometric properties of the faintest galaxies
in the EDisCS sample. The mean $I$ magnitude was set to be
$<I>=$24.00(24.5) for the low-z(high-z) cluster fields, corresponding
to a $S/N \sim 5$.  For all the other $j$ filters, magnitudes were
scaled according to the original SEDs, i.e. keeping colors unchanged:
$m_{new}(j)= m(j) + [24.0/24.5 - I]$
Photometric errors as a function of apparent magnitudes were
introduced and assigned as in \S~\ref{simus}. In this case, $\Delta
m_{new}^2(j) = [2.5 \log [1+1/(S/N)]]^2 - \Delta^2 m(j) $, where
$\Delta m(j)$ is the catalog error corresponding to $m(j)$, and $S/N =
S/N(m(j)) 10^{-0.4(m_{new}(j)-m(j))}$.  This procedure conserves
globally the colors of galaxies. The main caveat is the fact that this
noisy population does not necessarily match the true color
distribution of the faintest galaxies in the sample. However, it is
useful to estimate the degradation expected in \zphot accuracy between
the brightest and the faintest galaxies because of the lowered $S/N$.
Because \zphot quality is quite insensitive to spectroscopic quality,
we added the type 1 $+$ type 2 spectroscopic catalogs.

Table~\ref{tab_zzfaint} and ~\ref{GR_tab_zzfaint} summarize the results
obtained for the faintest galaxies using {\it Hyperz} and the GR code
respectively. These tables can be compared directly with Table~\ref{zzsum}.
The quality of photometric redshifts degrades typically by a factor of
two in $\sigma(\Delta z/(1+z))$ between the brightest ($I\ltapprox$22)
and the faintest ($I\sim$24-24.5) galaxies in the EDisCS sample. Most
of the trends observed in Table~\ref{zzsum} for the spectroscopic
sample are found in the 
Table~\ref{tab_zzfaint} for the simulations of the faintest sample, 
in particular
the lack of systematics in $\Delta_z$, and the higher quality results
in the high-z bin. The fraction of catastrophic identifications
increases, but remains typically below $\sim$ 5\%. The difference in
\zphot quality between early and late types is smaller for the
faintest galaxy sample. 
In this case, the simulation results with the GR code are found to be of
slightly higher accuracy than {\it Hyperz}'s ones (by $\ltapprox$20\% in
$\sigma(\Delta z/(1+z))$).
In Sect.~\ref{prob_maglim}, we comment on the
implications that this results will have for the calculation of
membership using the photometric redshifts.

%%%%%%%%%%%%%%%%%%%%%%%%%%%%%%%%%%%%%%%%%%%%%%%
%%%%%%%%%%%%%%%%%%%%%%%%%%%%%%%%%%%%%%%%%%%%%%%
\section{\label{cluster} Photometric determination of cluster redshifts}

\subsection{\label{preselection} Spectroscopic sample preselection}

Before the first spectroscopic runs, cluster redshifts were estimated
from the first \zphot catalogs using {\it Hyperz}. Spectroscopic
targets were selected mainly to have \zphot within the interval
z$_{cluster} \pm 0.2$, or
{\it absolute} P(z$_{cluster}$) $\ge 0.5$,
and according to the magnitude selection (see Halliday et al.\
2004).
Although the results discussed in this Sect. were obtained with {\it
  Hyperz}, they should be representative of the general behavior of
all SED-fitting \zphot codes.

\begin{figure*}[!ht]
{\centering \leavevmode
\psfig{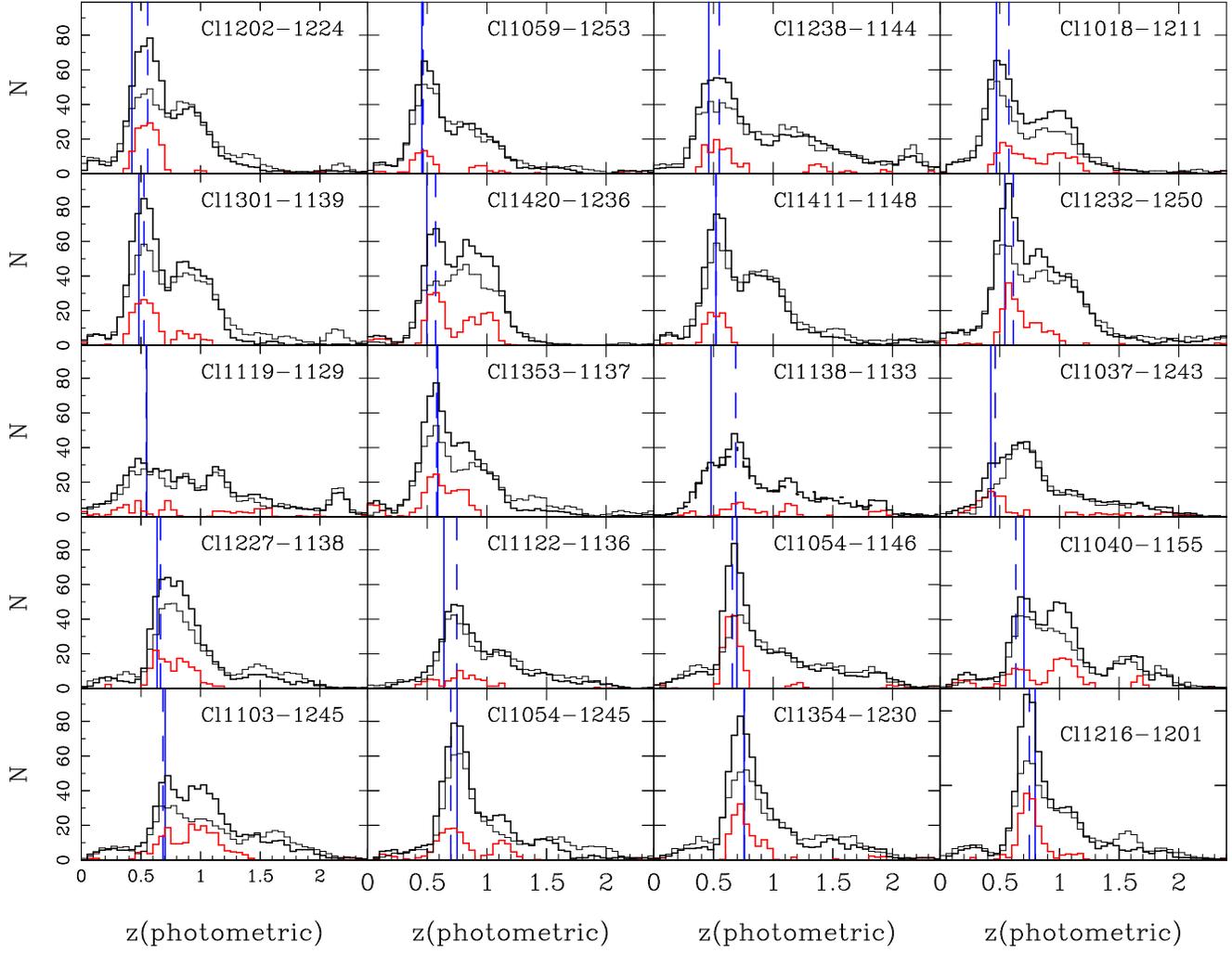}
}
\caption{Photometric redshift distributions in the EDisCS fields,
  with the cluster redshift increasing from top to bottom and from
  left to right, 
for the low-z and high-z samples (first and second series respectively).
The histograms display the following redshift
  distributions: 
N$_{in}$(z) (black thick solid lines), N$_{out}$(z) (thin
  black lines), and the positive difference between N$_{in}$(z) -
  N$_{out}$(z) (lowest histograms, thick red lines).
A real cluster or other structure
  corresponds to a positive excess in the lower (red) histogram.
  Histograms were obtained with a $\delta z =0.05$ sampling step,
  smoothed by a $\delta z =0.15$ sliding window. Vertical solid and
  dashed lines indicate the spectroscopic and photometric values,
  respectively, adopted for the cluster redshift. All vertical
  scales are identical in the number of galaxies / $\Delta z =$ 0.05,
  from N(z)$=$0 to 100, apart from Cl1216-1201, for which the range is 0 to 115.
}
\label{nz_1}
\end{figure*}

The photometic cluster redshifts were computed from the photometric
redshift distribution by comparing the N(z) obtained in the center of
the field with the equivalent one over a wider region of the same
area, obtained under the same conditions from the \zphot point of view
(same effective exposure time and number of filters), and used as a
blank field. A real cluster or other structure should have appeared as
an excess of galaxies in the central region in comparison to the outer
parts. In this exercise, we considered only objects with N$_G$ $\ge$ 1
i.e. objects that could not be excluded as galaxies without applying a
cut in magnitude. Figure~\ref{nz_1} displays the results found for the
different fields.The histograms in this Figure display the difference
between the redshift distribution within a $\sim$ 140'' radius region
centered on the center of the image (N$_{in}$(z), black solid line),
and the distribution within an outer ring, $140" \le r \le 200"$,
(N$_{out}$(z), dashed black lines). Red solid lines show the positive
difference between the two histograms, N$_{in}$(z) - N$_{out}$(z).
Histograms were obtained with a $\delta z =0.05$ sampling step and
smoothed with a $\delta z =0.15$ sliding window. This window
corresponded approximately to the $1\sigma$ uncertainty in the \zphot
estimate for the faintest galaxies in the catalog.

Where there was a distinct peak in N$_{in}$(z) - N$_{out}$(z)
distribution, we used this value to represent the "cluster
redshift". We also computed 2D number density maps and cluster
tomography (see \S~\ref{tomography} below) to emphasize the reality of
the clusters, in particular for the uncertain cases. A summary of
these results was provided in White et al.\ (2005). We note that the
efficiency of the cluster-finding algorithm could be enhanced if the
central region was centered on the cluster centroid instead of the
center of the image. This ideal situation could be achieved in wider
surveys.
 
%%%%%%%%%%%%%%%%%%%%%%%%%%%%%%%%% Fig. cluster histos & tomography

%%%%%%%%%%%%%%%%%%%%%%%%%%%%%%%%% end Fig. cluster histos & tomography

%%%%%%%%%%%%%%%%%%%%%%%%%%%%%%%%%%%%%%%%%%%%%%%
\subsection{Spectroscopic versus photometric cluster redshifts}

Figure~\ref{nz_1} summarizes the comparison between the \zphot and
spectroscopic redshifts for the different cluster fields.
The photometric cluster redshift can be defined in different
ways. Here we have adopted two different definitions, which are
reported in Table~\ref{zzcluster}: (1) the mean weighted value,
computed from the excess peak, and (2) the redshift corresponding to
the maximum value in the N$_{in}$(z) - N$_{out}$(z) histogram.
Table~\ref{zzcluster} provides a comparison between spectroscopic and
photometric determinations of cluster redshifts for the low and high-z
samples. Cluster redshifts in Table~\ref{zzcluster} and
Fig.~\ref{nz_1} correspond to 
the most prominent cluster identification
when several clusters were present in the field (Milvang-Jensen et al.\
2008). Five fields were excluded when computing the systematic
deviation and dispersion: Cl1119-1129 and Cl1238-1144 in the low-z
sample, because of incomplete photometry; and Cl1122-1136,
Cl1037-1243a and Cl1138-1133 
in the high-z sample, the first because of the
lack of a clear cluster in the field, and the two others because their
low cluster redshifts implied that B-band photometry was required to
achieve an accurate \zphot (see also Milvang-Jensen et al.\ 2008). We
note however the accurate photometric identification of the Cl1037-1243a cluster,
for which a poor determination was expected.

In general, the differences between photometric and spectroscopic
values are small, ranging from $\delta z \sim$0.03-0.04 at high-z to
$\delta z \sim$0.05 at low-z. The dispersion is much lower than the
approximate cut introduced by the spectroscopic preselection
(z$_{cluster} \pm 0.2$). Therefore, there is no reason why a cluster
should have been ``missed'' within the relevant redshift interval due
to \zphot preselection. The systematic trend of lower quality in
\zphot for the low-z sample was expected from the simulations
presented in \S~\ref{simus}, and also observed in the comparison with
the spectroscopic sample in \S~\ref{zzspectro}.

%% TABLE z(cluster) spectro vs zphot (SMOOTHED VERSION)

\begin{table}
\begin{center}
\caption{\label{zzcluster}Comparison between spectroscopic (z$_{cl}$) and photometric
  determinations of cluster redshifts in the low and high-z
  samples.  }
\begin{tabular}{cccccc}
\hline \hline
Cluster (low-z) & z$_{cl}$ & \zphot & z$_{cl}$ -
z$_{phot}$ & \zphot  & z$_{cl}$ -z$_{phot}$ \\
   &              & (1) & (1) & (2)& (2) \\
\hline 
Cl1018-1211    & 0.473 & 0.575 & -0.102 & 0.525 & -0.052 \\
Cl1059-1253    & 0.456 & 0.465 & -0.009 & 0.478 & -0.022 \\
Cl1119-1129$^1$    & 0.550 & 0.544 &  0.006 & 0.475 & 0.075 \\
Cl1202-1224    & 0.424 & 0.556 & -0.132 & 0.568 & -0.144 \\
Cl1232-1250    & 0.541 & 0.614 & -0.073 & 0.575 & -0.034 \\
Cl1238-1144$^1$    & 0.460 & 0.548 & -0.088 & 0.521 & -0.061 \\
Cl1301-1139    & 0.482 & 0.525 & -0.043 & 0.524 &  -0.042 \\
Cl1353-1137    & 0.588 & 0.579 &  0.009 & 0.573 &  0.015 \\
Cl1411-1148    & 0.520 & 0.520 &  0.000 & 0.568 & -0.048 \\
Cl1420-1236    & 0.496 & 0.570 & -0.074 & 0.569 & -0.073 \\
\hline
$\left<\Delta_z \right>$ &       &      & -0.053 &  
& -0.050 \\
  &       &      & $\pm$  0.050 &  
& $\pm$ 0.046 \\
$\left<|\Delta_z|\right>$ &       &      & 0.055 &  
& 0.053  \\
 &       &      & $\pm$ 0.048 &  
&  $\pm$ 0.041  \\
Median $\Delta_z$ &       &      & -0.058  &   & -0.045 \\
\hline \hline
Cluster (high-z) & z$_{cl}$ & \zphot  & z$_{cl}$ -
z$_{phot}$ & \zphot  & z$_{cl}$ -z$_{phot}$ \\
   &              & (1) & (1) & (2)& (2) \\
\hline 
Cl1037-1243a$^1$    &  0.425 & 0.461 & -0.036 & 0.424 & 0.001 \\
Cl1040-1155   &  0.704 & 0.635 &  0.069 & 0.624 & 0.080 \\
Cl1054-1146   & 0.697  & 0.658 &  0.039 & 0.673 & 0.024  \\
Cl1054-1245   & 0.750  & 0.697 &  0.053 & 0.727 & 0.023  \\
Cl1103-1245b   & 0.703  & 0.685 & 0.018 & 0.725 & -0.022 \\
Cl1122-1136$^1$   & 0.640  & 0.748 & -0.108 & 0.773 & -0.133 \\
Cl1138-1133$^1$   & 0.479  & 0.686 & -0.207 & 0.720  & -0.241 \\
Cl1216-1201   & 0.794  & 0.747 &  0.047 & 0.725 & 0.069 \\
Cl1227-1138    & 0.635 & 0.664 & -0.029 & 0.625 & 0.010 \\
Cl1354-1230   &  0.762 & 0.759 &  0.003 &  0.724 & 0.038 \\
\hline
$\left<\Delta_z \right>$ &       &      & 0.028  &  
& 0.031  \\
 &       &      &  $\pm$  0.033 &  
&  $\pm$ 0.035 \\
$\left<|\Delta_z|\right>$ &       &      & 0.037  &
& 0.038 \\
 &       &      & $\pm$ 0.022 &
&  $\pm$ 0.026\\
Median $\Delta_z$ &       &      & 0.039 &    &  0.024 \\
\hline 
\end{tabular}
\end{center}

Notes -- Two estimates for the photometric determination are
  provided: (1) mean weighted value, and (2) the redshift
  corresponding to the maximum of the N(z) distribution. Clusters
  labeled with $^1$ were excluded when computing the systematic
  deviation and dispersion (see text).
\end{table}

%%%%%%%%%%%%%%%%%%%%%%%%%%%%%%%%%%%%%%%%%%%%%%% fig tomography %%%%%% 

\begin{figure*}[!ht]
\psfig{file=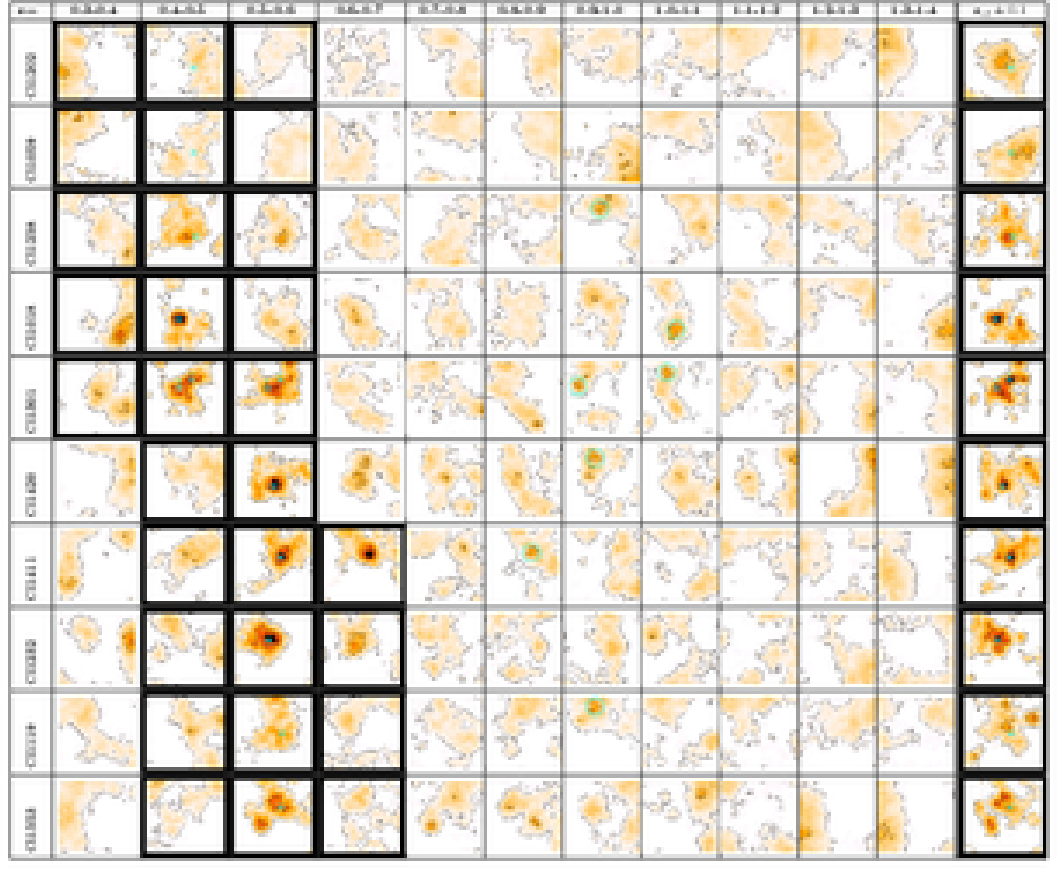,angle=270,width=1.0\textwidth}
%\includegraphics[angle=270,width=1.0\textwidth]{figures_papier/tomography_lowz.pdf}
%}
\caption{Projected number density maps for the low-$z$ sample, for
  different redshift slices with z$_{cluster}$ increasing from top to
  bottom. Projected number densities are displayed in a linear scale.
  Isopleths correspond to increasing number density bins with $\Delta
  \Sigma_{20}$ = $<\Sigma_{20}>$, starting at the mean $\Sigma_{20}$
  within the redshift slice. Thick frames highlight the redshift
  slices encompassing z$_{cluster} \pm 0.1$.  The rightmost column
  displays the density map for the z$_{cluster} \pm 0.1$ redshift
  slice, where the maximum contrast in the density peak is usually
  reached. 
The position of the BCG for the most prominent clusters in the field are
displayed by a blue cross. For
  Cl1059-1253, Cl1202-1224, and Cl1119-1129, only two isopleths are
  displayed corresponding to 1 and 1.5 $<\Sigma_{20}>$.  Additional
  overdensities along the line of sight with detection levels
  exceeding $4\sigma$ are displayed by circles.  
{\bf High resolution figure available at
http://www.ast.obs-mip.fr/users/roser/photoz\_EDISCS/10644f8.pdf}
}
\label{tomo_lowz}
\end{figure*}

\begin{figure*}[!ht]
\psfig{file=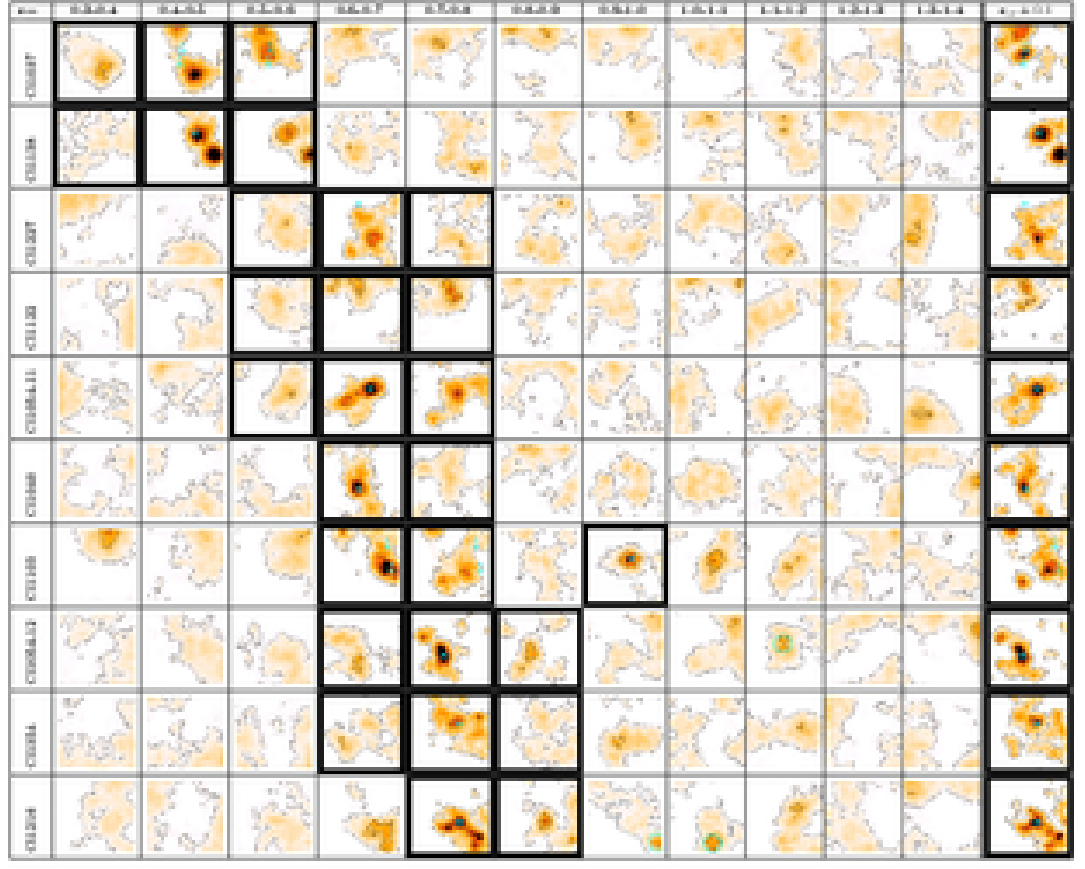,angle=270,width=1.0\textwidth}
\caption{Projected number density maps for the high-$z$ sample, for
  different redshift slices. Same comments as in Fig.~\ref{tomo_lowz}.
  There is no clear BCG identified in Cl1122-1136.  
{\bf High resolution figure available at
http://www.ast.obs-mip.fr/users/roser/photoz\_EDISCS/10644f9.pdf}
}
\label{tomo_highz}
\end{figure*}

%\end{landscape}

%%%%%%%%%%%%%%%%%%%%%%%%%%%%%%%%%%%%%%%%%%%%%%%
%%%%%%%%%%%%%%%%%%%%%%%%%%%%%%%%%%%%%%%%%%%%%%%
\subsection{\label{tomography} Cluster tomography}

We present results about cluster tomography along the line-of-sight in
the different EDisCS fields. This highlights the capability of \zphot
in identifying and studying clusters of galaxies in deep photometric
surveys.

A local density estimate was derived at each point in the field
using a grid with
$\delta x= \delta y = $1''. The density estimator was defined,
according to the formalism introduced by Dressler (1980), to equal
$\Sigma_{20}=20/(\pi d_{20}^2)$, where $d_{20}$ was the projected
linear distance to the 20-th closest neighbor. Photometric redshift
slices of $\Delta z=$ 0.1 were used to cover the redshift range $0\le
z \le 1.4$. Close neighbors were selected within \zphot$\pm0.1$,
centered on the redshift bin. The arbitrary choices of $\Sigma_n$ and
n$=$20 were justified by the typical richnesses of clusters and the
\zphot accuracy. We tested other values of n ranging between 5 and 20
and achieved similar results for the cluster detection. Edge effects
were corrected by using external fields in the main eight directions
(i.e. X+, Y+, X-, Y-, X+Y+, X+Y-, X-Y+, X-Y-),
containing different realizations of the same \zphot slice with xy
coordinates randomly sorted.  The density maps were smoothed with a
Gaussian kernel of $\sigma=$5".
    
Figures~\ref{tomo_lowz} and ~\ref{tomo_highz} display the projected
number density maps obtained with this method, for the low and high-z
samples, respectively. In these figures, clusters are presented with
z$_{cluster}$ increasing from top to bottom. Projected number
densities are displayed on a linear scale, for an arbitrary redshift
step $\Delta z=0.1$, with isopleths corresponding to increasing number
density bins, equally spaced 
with $\Delta \Sigma_{20}$= $<\Sigma_{20}>$, starting at
the mean $\Sigma_{20}$ within the redshift slice. The presence of a
cluster along the line-of-sight is clearly seen in most cases, at
least in all cases where a cluster was clearly detected in the
field. We note that the $<\Sigma_{20}>$ value is affected by the
presence of a cluster in the corresponding redshift bin. The most
significant contrast in the density map is found usually for the
z$_{cluster} \pm 0.1$ redshift slice (rightmost column of
Fig.~\ref{tomo_lowz} and ~\ref{tomo_highz}). The position of the BCG
(White et al.\ 2005; Milvang-Jensen et al.\ 2008; 
Whiley et al.\ 2008; 
displayed by a blue cross in these Figures)
coincides usually with the maximum contrast in the density map
for the most prominent clusters in a given field. The
comparison between this maximum-contrast slice and the tomography with
fixed arbitrary step (e.g. $\Delta z=0.1$, the step used in
Fig.~\ref{tomo_lowz} and ~\ref{tomo_highz}) suggests that the optimal
redshift step for detection of clusters along the line-of-sight should
be close to the typical difference between photometric and
spectroscopic z$_{cluster}$, in this case $\Delta z \sim
0.05$. Structures separated by less than $\Delta z \sim 0.05$ in
redshift space cannot be distinguished by \zphot tomography
(e.g. clusters Cl1138 (z=0.48) and Cl1138a (z=0.45) in the 1138.2-1133 field,
  and Cl1227 (z=0.63) and Cl1227a (z=0.58) in the 1227.9-1138 field;
Milvang-Jensen et al.\ 2008).

For all clusters clearly identifiable in this sample (i.e. all
clusters except Cl1059-1253, Cl1202-1224 and Cl1119-1129), the maximum
contrast is found to be at least $\sim 3-4 \times <\Sigma_{20}>$ about
z$_{cluster} \pm 0.1$. All clusters in our sample exhibit a
significant $\gtrsim 3\sigma$ overdensity around z$_{cluster}\pm0.1$,
defined to be $\Sigma_{20}-<\Sigma_{20}> / \sigma(z)$, where
$\sigma(z)$ represents the standard deviation in the projected number
density within the redshift bin.  For all apart from the three
aforementioned clusters, the detection level exceeds $4\sigma$, and
ranges between 6 and $9\sigma$ for Cl1216-1201, Cl1227-1138,
Cl1411-1148, Cl1420-1236, Cl1040-1155, Cl1054-1245, Cl1138-1133,
Cl1018-1211, and Cl1054-1146.
%% Other significant structures?
No other significant overdensities are found along the line-of-sight
of density peaks exceeding $3 \times <\Sigma_{20}>$, which are typical values
for rich clusters in this sample. However, several overdensities are
found with slightly smaller values, around $\sim 2-3 \times <\Sigma_{20}>$,
and detection levels exceeding $4\sigma$. These structures are
identified by circles in Figures ~\ref{tomo_lowz} and
~\ref{tomo_highz} (see below).  Their reliability is difficult to
assess with the presently available data.

Several fields deserve further comment. More information concerning
the spectroscopic identification of clusters in these fields can be
found in the reference papers by Halliday et al.\ (2004) and
Milvang-Jensen et al.\ (2008):

\begin{itemize}

\item Cl1301-1139: The two clusters identified in this field with
  z(spectroscopic)=0.397 (Cl1301a) 
and 0.482 (Cl1301) are consistent with the two
  different $\gtrsim 3 <\Sigma_{20}>$ peaks observed by our
  tomography.

\item Cl1037-1243: The most distinct $\sim 3 <\Sigma_{20}>$
  overdensity in this field corresponds to the ``a'' component at
  z(spectroscopic)=0.425, whereas the first identification was given
  at z(spectroscopic)=0.578 (z$_{cluster} \pm 0.1$ in
  Fig. ~\ref{tomo_highz}). Both structures are seen by tomography.

\item Cl1354-1230: the two clusters Cl1354 and Cl1354a 
 ($z=$0.762 and
  $z=$0.595 respectively) are consistent with overdensities at the
  $\gtrsim 2 <\Sigma_{20}>$ level.

\item Cl1103-1245: The prominent 
 cluster detected in this field at
  $z=0.96$ is clearly visible ($\gtrsim 8\sigma$ level) both in the
  density map and the \zphot distribution (Fig.~\ref{nz_1}). 
The two components Cl1103a ($z=$0.626) and Cl1103b ($z=$0.703) 
can hardly be separated by tomography.

\item Other secondary peaks at z$\sim 0.9-1.1$ are seen in the
  N(\zphot) distribution of Cl1018-1211, Cl1301-1139, Cl1420-1236,
  Cl1054-1245, and Cl1040-1155 (Fig.~\ref{nz_1}). All can be
  associated with spatial overdensities of $2-3 <\Sigma_{20}>$ at the
  $4-5\sigma$ level, apart from in the field of Cl1040-1155, where no
  significant overdensity is found.

\item Several additional overdensities/structures at z$\sim 0.8-1.0$
  are found in Cl1411-1148, Cl1119-1129, Cl1238-1144, and Cl1216-1201,
  with $2-3 <\Sigma_{20}>$ and detection levels ranging between 4 and
  6 $\sigma$.  Given the limited photometric coverage of Cl1119-1129
  and Cl1238-1144, the detected overdensities in these fields are
  rather dubious.

\end{itemize}

%%%%%%%%%%%%%%%%%%%%%%%%%%%%%%%%%%%%%%%%%%%%%%%
%%%%%%%%%%%%%%%%%%%%%%%%%%%%%%%%%%%%%%%%%%%%%%%
\section{\label{member} Cluster membership criteria}

The most unambiguous way to determine cluster membership is by means
of accurate spectroscopic redshifts. Unfortunately, it is far too
time-consuming to obtain high spectroscopic completeness in cluster
member observations, even to relatively bright limits of $I< 22-23$.
For this reason, it is necessary to develop membership criteria that
rely solely on photometric data. To achieve many EDisCS science goals,
such as study of luminosity functions and cluster substructure, any
method should: 1) retain $>$90\% of cluster members, 2) reject an
optimal number of non-members, and 3) measure a probability that a
given galaxy is a cluster member.  The first two criteria should be
implemented so that there is little dependence on the galaxy color,
e.g. for Butcher-Oemler-type studies. While traditional methods of
statistical subtraction using ``field'' surveys of comparable depth
offer a viable method to satisfy the first two criteria, they do not
satisfy the third. For this reason, we developed an alternative method
for membership determination based on our photometric redshifts
estimates.

We present below
how we use the photometric redshift probability
distribution \pz to reject non-members from each cluster field. We
describe the method that we developed and its calibration based on
EDisCS spectroscopic redshifts. We discuss how this method can be
extended to the entire magnitude-limited sample for a given cluster
and outline its limitations.

%%%%%%%%%%%%%%%%%%%%%%%%%%%%%%%%%%%%%%%%%%%%%%%
\subsection{\label{prob_method} The method}

Traditionally, \zp--based methods for determining cluster membership
were based on a simple cut in redshift, such that a galaxy was
considered to be a member if $|z_{phot} - z_{clust}| < \Delta
z_{thresh}$. One disadvantage of this method is that $\Delta
z_{thresh}$ can be as high as 0.3 (e.g. Toft et al. 2004), causing
considerable field contamination to enter into the cluster sample. An
additional disadvantage of the method is that it uses only the
best-fit redshift in determining membership and ignores the
information contained in the full redshift probability distribution
\pz.

\begin{figure}[!h]
%{\centering \leavevmode
\psfig{file=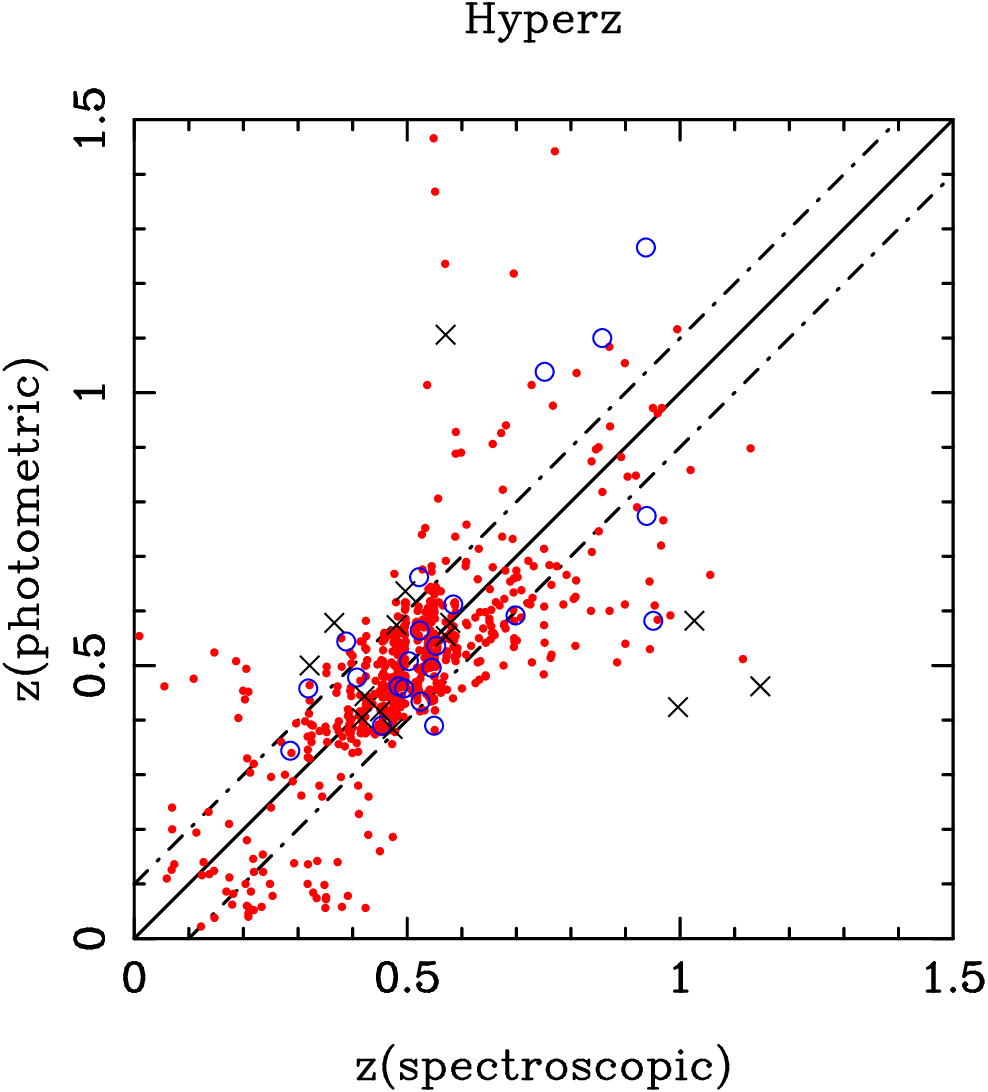,angle=270,width=0.43\textwidth}
\newline
\psfig{file=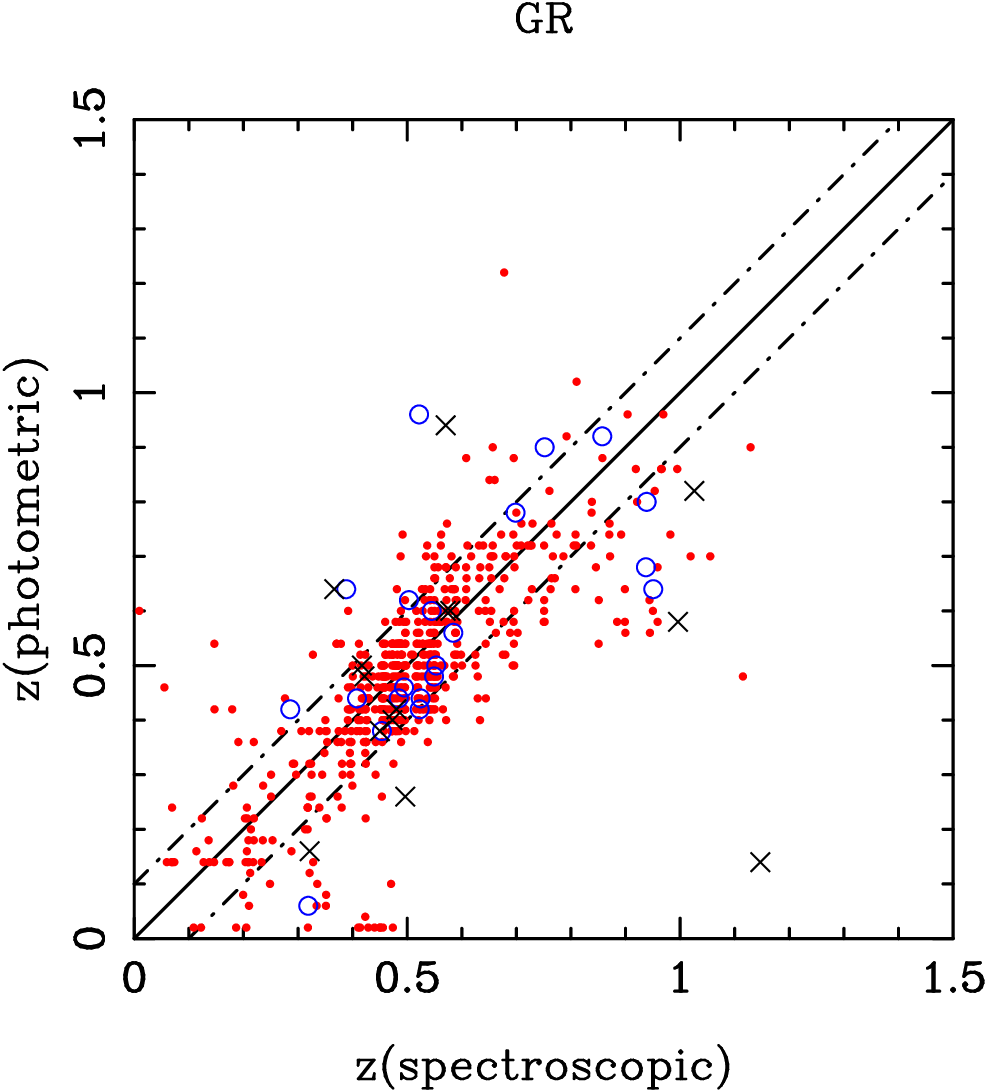,angle=270,width=0.43\textwidth}
\newline
\psfig{file=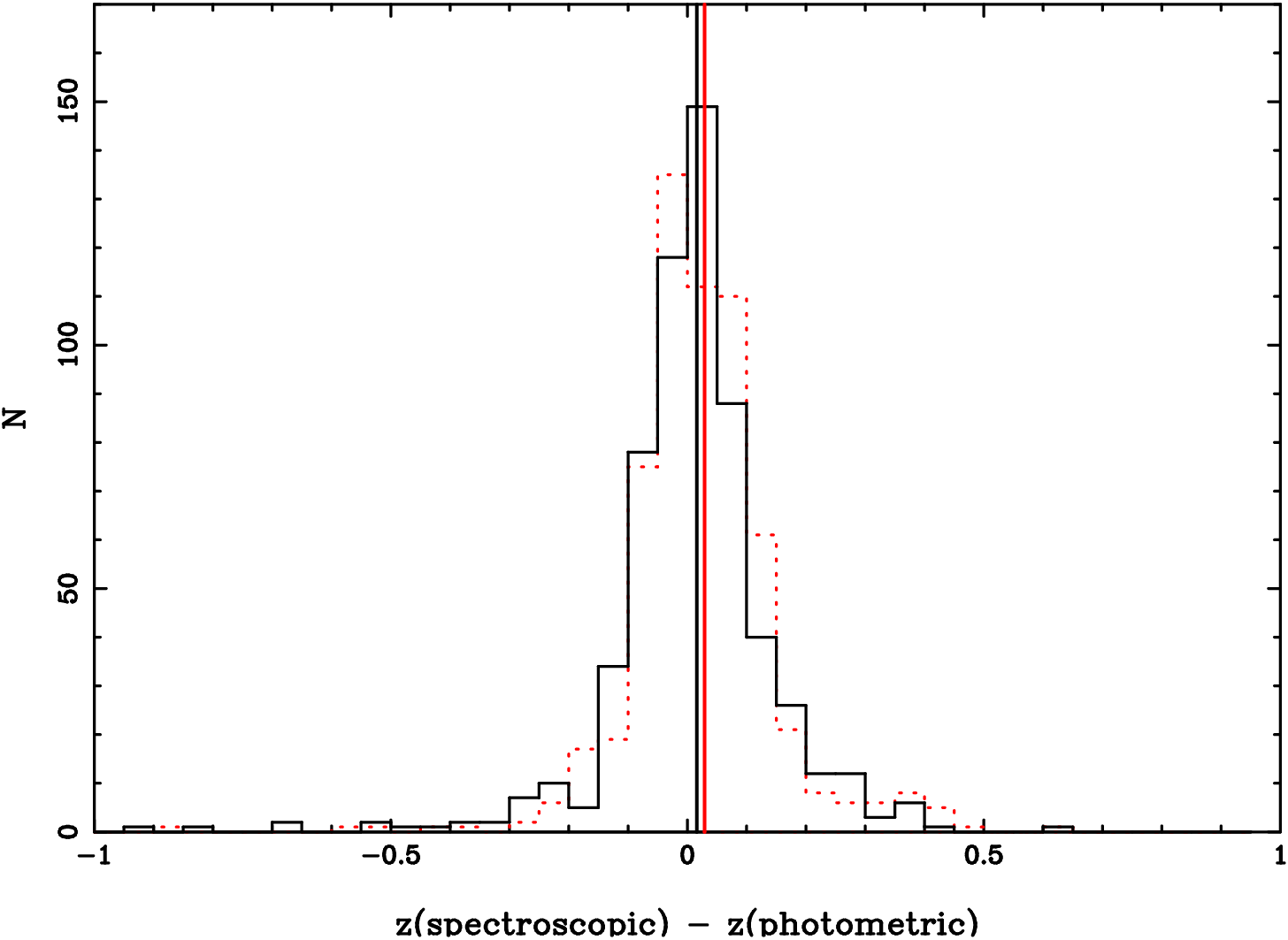,angle=270,width=.50\textwidth}
\caption{Comparison between spectroscopic and photometric redshifts
  for the low-$z$ sample, obtained with {\it Hyperz} (top panel) and
  GR (central panel) codes. Solid (red) circles, open (blue) circles,
  and crosses correspond to objects with good (type 1), medium (type
  2) and tentative (type 3) spectroscopic redshift determinations,
  respectively. 
Dot-dashed lines display \zspec $=$  \zphot $\pm 0.1$ to guide the eye.
The bottom panel displays the $z_{\rm spec} - z_{\rm
    phot}$ distribution obtained for this sample with {\it Hyperz}
  (solid black line) and GR (dotted red line). Vertical lines indicate
  the mean $\left<\Delta_z \right>$.  
\label{zz_low}
}
\end{figure}

\begin{figure}[!h]
%{\centering \leavevmode
\psfig{file=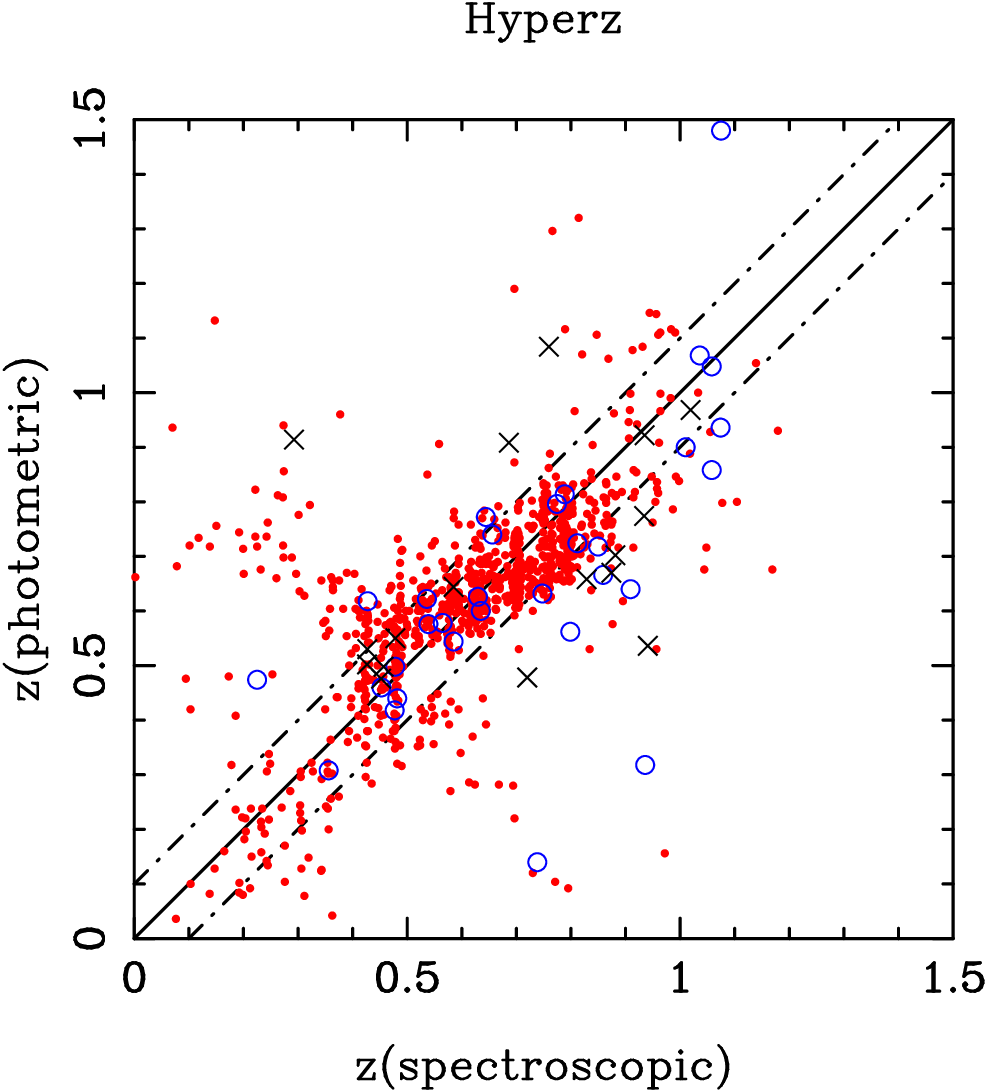,angle=270,width=0.43\textwidth}
\newline
\psfig{file=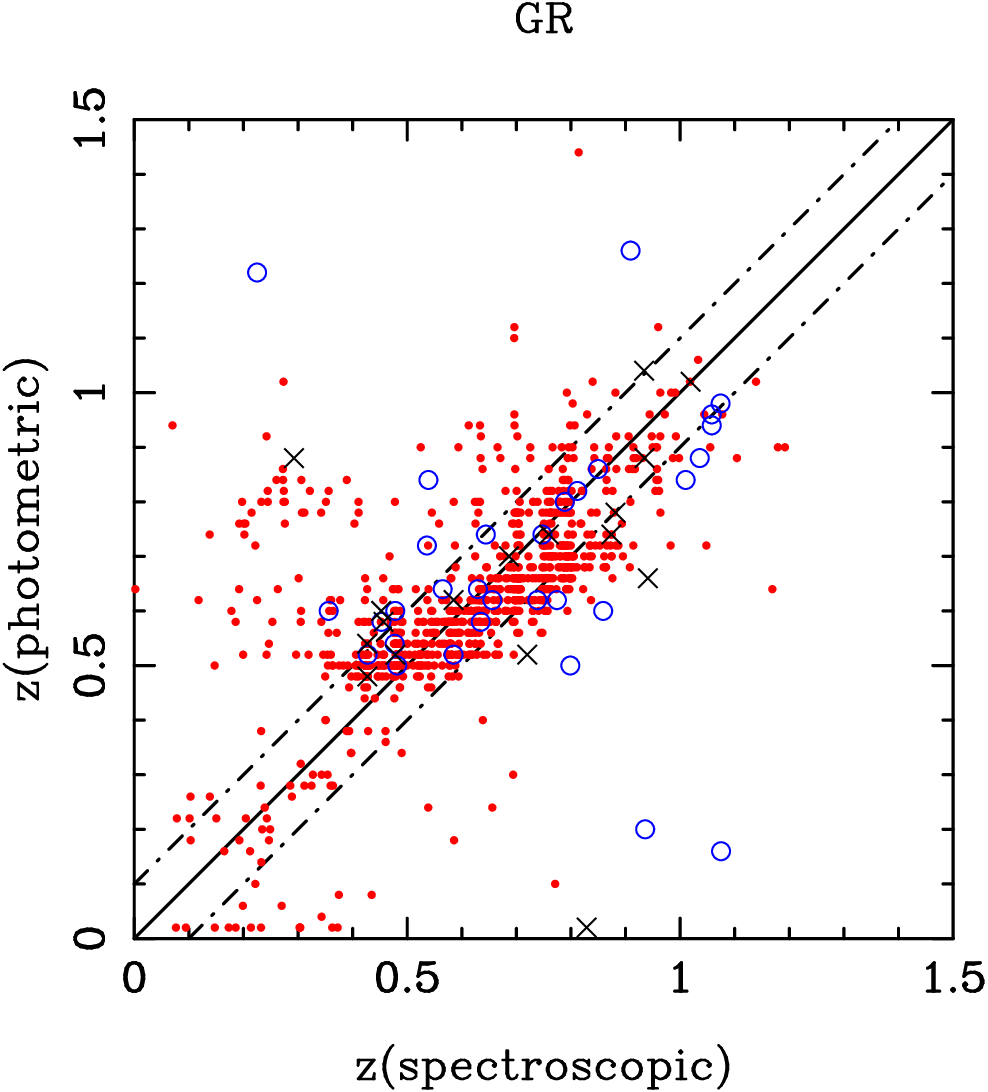,angle=270,width=0.43\textwidth}
\newline
\psfig{file=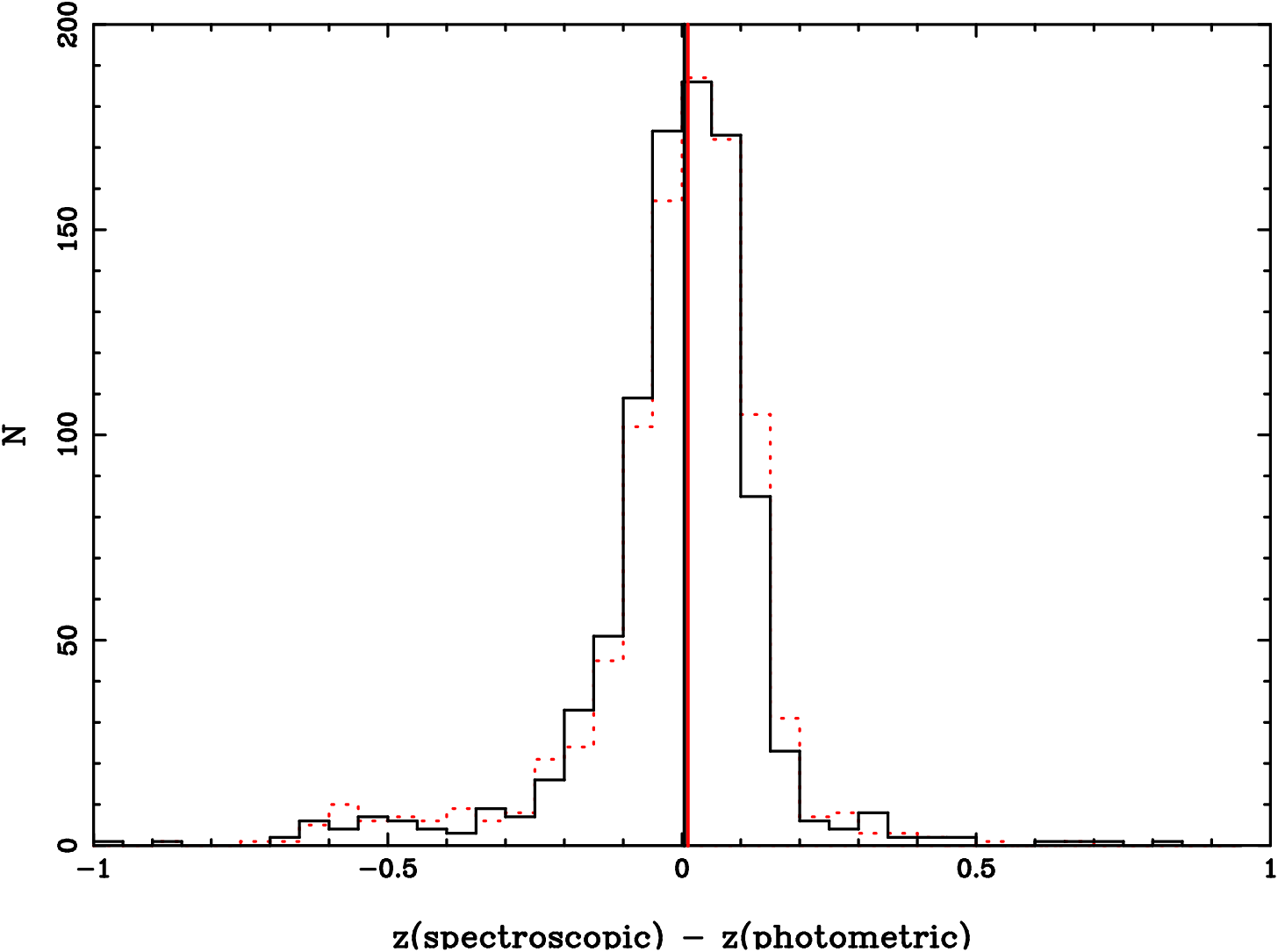,angle=270,width=.50\textwidth}
\caption{(Comparison between spectroscopic and photometric redshifts
  for the high-$z$ sample, obtained with {\it Hyperz} (top panel) and
  GR (central panel) codes. Solid (red) circles, open (blue) circles,
  and crosses correspond to objects with good (type 1), medium (type
  2) and tentative (type 3) spectroscopic redshift determinations,
  respectively. 
Dot-dashed lines display \zspec $=$  \zphot $\pm 0.1$ to guide the eye.
The bottom panel displays the $z_{\rm spec} - z_{\rm
    phot}$ distribution obtained for this sample with {\it Hyperz}
  (solid black line) and GR (dotted red line). Vertical lines indicate
  the mean $\left<\Delta_z \right>$.  
\label{zz_high}
}
\end{figure}

Brunner \& Lubin (2000) suggested an improved technique that used \pz
in determining galaxy membership. They assumed a Gaussian \pz of width
calculated from the comparison with \zs, and defined the quantity

\begin{equation}
P_{clust} = \int^{z_{clust}+\Delta z/2}_{z_{clust}-\Delta z/2}P(z)dz,
\end{equation}

where $\Delta z$ was an interval defined around \zclust that reflected
the dispersion in the \zp versus \zs diagram, and \zclust was the
spectroscopic redshift of the cluster. 
In Section ~\ref{prob_specsamp} we explain in detail the method we used
to calibrate a threshold value for \pclust, \pthresh, below which a galaxy
would be considered to be a non-member. 
However, in reality, \pz can be highly non-Gaussian,
with multiple maxima and extended tails at large distances from the
most likely solution. For this reason, it may not be optimal to assume
the Gaussian approximation. We therefore extended the Brunner \& Lubin
method to use the full \pz dataset calculated directly from the two
\zp codes. We tested the accuracy of our \pz by comparing the
confidence intervals derived from \pz with the disagreement between
\zs and \zp (see Figures~\ref{zz_lowz1} to ~\ref{zz_high} in
\S~\ref{dz}).
The $z_{spec}$ fell within the 68\% confidence limits on $z_{phot}$
for $\sim 68\%$ of the galaxies and many of the galaxies with large
$|z_{phot}-z_{spec}|$ also had correspondingly large 68\% confidence
intervals.  This gave us assurance that $P(z)$ was accurate enough for our
purposes. 

%%%%%%%%%%%%%%%%%%%%%%%%%%%%%%%%%%%%%%%%%%%%%%%
\subsection{\label{prob_specsamp} Calibrating from the spectroscopic sample}

We adopt the large and uniform EDisCS spectroscopic sample (Halliday
et al.\ 2004; Milvang-Jensen et al.\ 2008) to calibrate the \pthresh
that we use to reject non-members.

\begin{figure}
\psfig{file=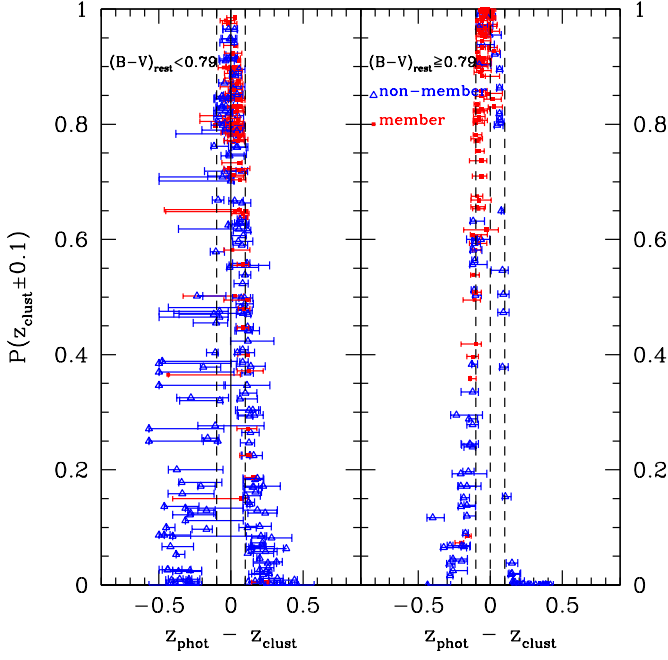,width=0.5\textwidth}
\caption { The integrated probability of being at \zclust$\pm0.1$
  vs. \zp-\zclust for the low-z clusters. All galaxies with secure
  spectroscopic redshifts were included. The different panels
  represent the blue and red halves of the sample in \bvr. The
  typically lower values of blue galaxies is due to their broader
  probability distributions. The solid vertical line indicates
  \zp=\zclust and the dashed lines indicate $\pm0.1$ in redshift.
  Horizontal error bars correspond to 68\% confidence intervals. }
\label{probz_lowz_restcol}
\end{figure}

\begin{figure}
\psfig{file=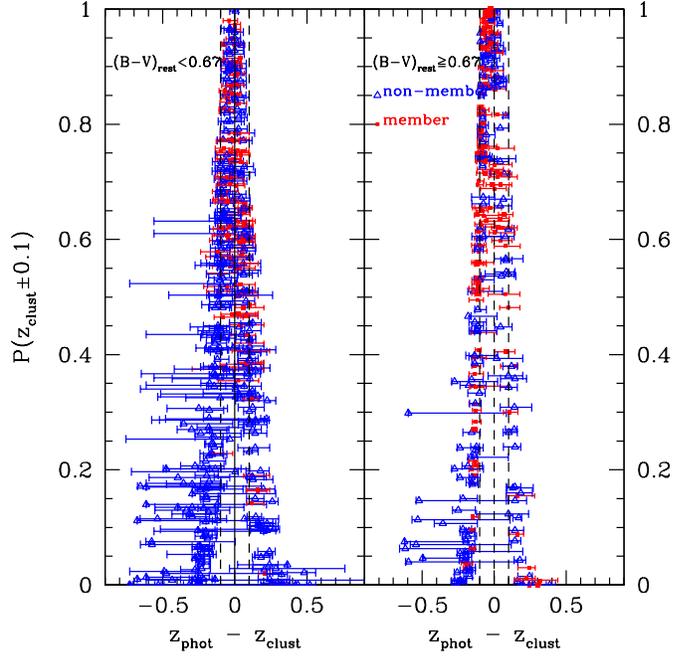,width=0.5\textwidth}
\caption { Same as Figure~\ref{probz_lowz_restcol} but for the high-z
  clusters.}
\label{probz_highz_restcol}
\end{figure}

We show in Figures~\ref{probz_lowz_restcol} and
\ref{probz_highz_restcol} the \pclust versus \zp-\zclust for all
galaxies with secure \zs measurements in our clusters, for the low and
high redshift samples, respectively. Results presented here were
obtained for the GR code, but they are similar for {\it Hyperz}. 
As seen in Section ~\ref{dz}, the \zp accuracy as well as the fraction of
catastrophic identifications in a given field depend on the spectral type of
galaxies, although the difference between early and late types is smaller for
the faintest galaxies in our sample. Therefore, we have studied the
reliability of the membership criteria for both early and late SED types
using a cut in the restframe color which splits the sample into two equal
halves of red (early) and blue (late) type galaxies. The color cuts are found to
be \bvr$=$0.79 and 0.67 for the low-z and high-z samples respectively. 
The left and right-hand panels in Figures~\ref{probz_lowz_restcol} and
\ref{probz_highz_restcol} provide results for objects of different spectral
types. 

We note that the ratio of members to non-members
increases as a function of \pclust. There are also few galaxies with
\zp$\approx$\zs and very low \pclust values. This implies that there
are not many members that would be rejected because their faint
magnitudes correspond to broad \pz and lead to their rejection even if
\zp$\approx$\zs. It is important to consider, however, that the
spectroscopic sample consists of the brightest galaxies of probably
the tightest \pz values and that this behavior might not be similar at
fainter magnitudes (see \S~\ref{prob_maglim}).

\begin{figure}
\psfig{file=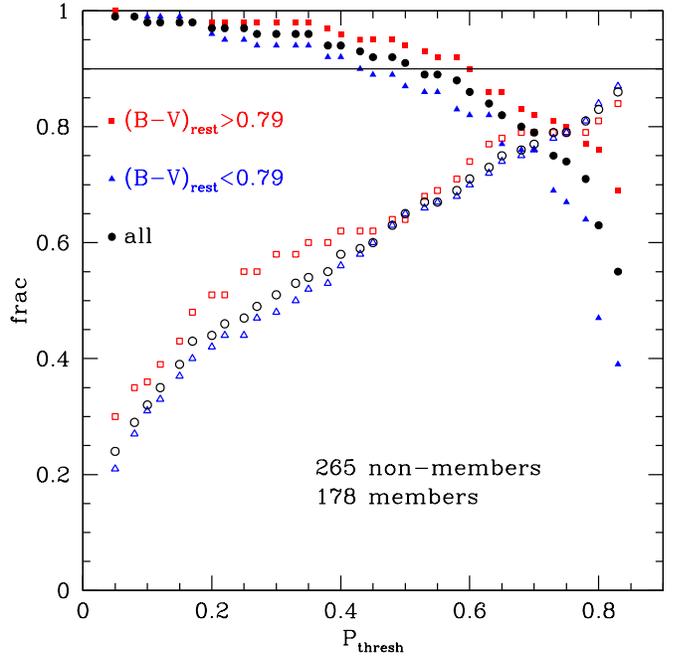,width=0.5\textwidth}
\caption {The fraction of retained members and rejected non-members as
  a function of the \pthresh and as a function of \bvr. This plot was
  created using all galaxies with secure \zs measurements 
in clusters with BVIK photometry.
  clusters. Open symbols represent the fractions of rejected
  non-members, while solid points are the fractions of retained
  members. The color and shape of the points indicate the \bvr cut
  applied. The solid horizontal line at 0.9 is included to guide the
  eye. 
This figure only shows the results for the GR code, as an
illustration of the technique.  It is not directly comparable to the
numbers quoted in Tables ~\ref{rejthreshtab} and ~\ref{rejfracspectab}, 
which utilize the combination of
both the GR code and {\it Hyperz}.
}
\label{pint_frac_lowz}
\end{figure}

\begin{figure}
\psfig{file=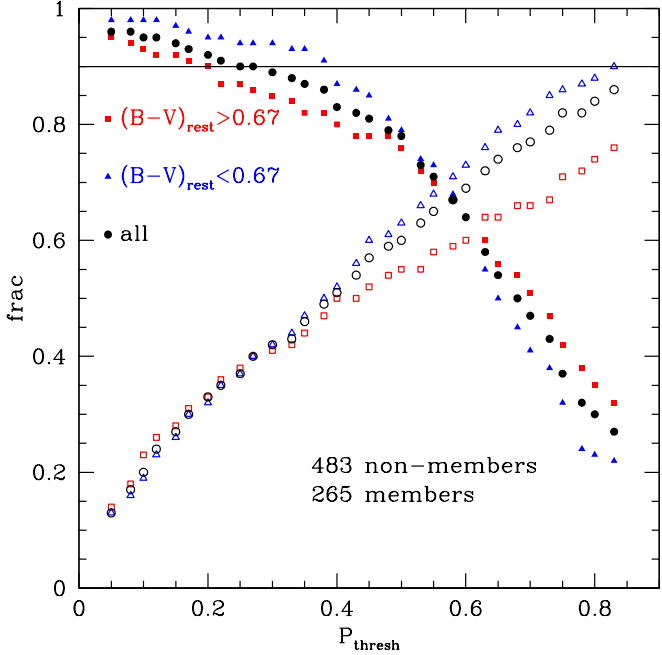,width=0.5\textwidth}
\caption {
Same as Figure~\ref{pint_frac_lowz} but for 
clusters with VRIJK photometry.
}
\label{pint_frac_highz}
\end{figure}

For the GR code, Figures~\ref{pint_frac_lowz} and
\ref{pint_frac_highz} demonstrate how the retained fraction of members
and rejected fraction of non-members depends on the \pclust threshold
\pthresh for the low-z and high-z samples, respectively.  In both
cases, it is possible to define a \pthresh value such that $>$90\% of
confirmed cluster members are retained with little dependence on
rest-frame (or observed) color.

\begin{figure*}[!h]
%{\centering \leavevmode
\psfig{file=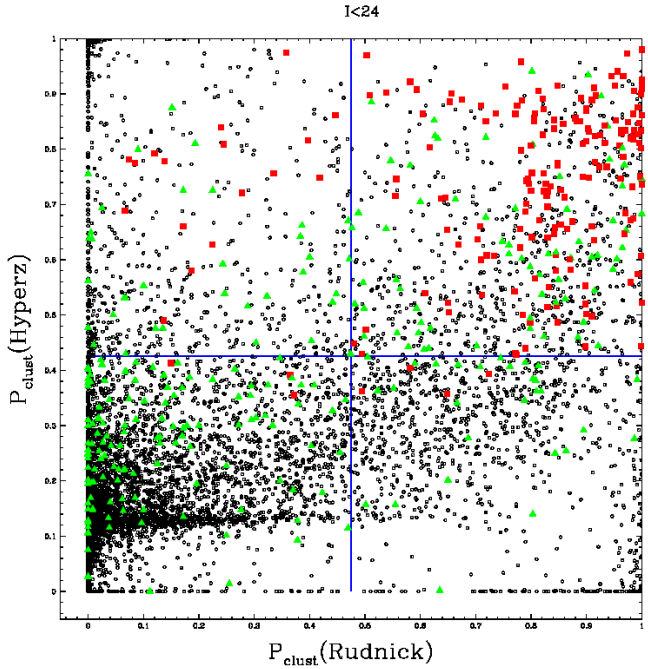,width=0.50\textwidth}
\caption {
A comparison of \pclust computed by the two different
  photometric redshift codes for the clusters with BVIK photometry.
  The black open
  circles represent all galaxies with $I<24.5$. The solid red squares
  are spectroscopically confirmed members and the solid green
  triangles are spectroscopically confirmed non-members. There is a
  broad correlation of \pclust between the two codes, such that the
  overwhelming number of galaxies have low \pclust for both methods
  with a smooth distribution of higher \pclust galaxies extending to
  the upper right, coincident with the confirmed members. The blue
  lines indicate the \pthresh values for each code that were
  determined to reject jointly the largest number of non-members,
  while retaining at least $90\%$ of the confirmed members. Objects to
  the left of the vertical blue line or beneath the horizontal blue
  line are flagged as interlopers.
\label{pint_comp_lowz}
}
\end{figure*}

\begin{figure*}[!h]
\psfig{file=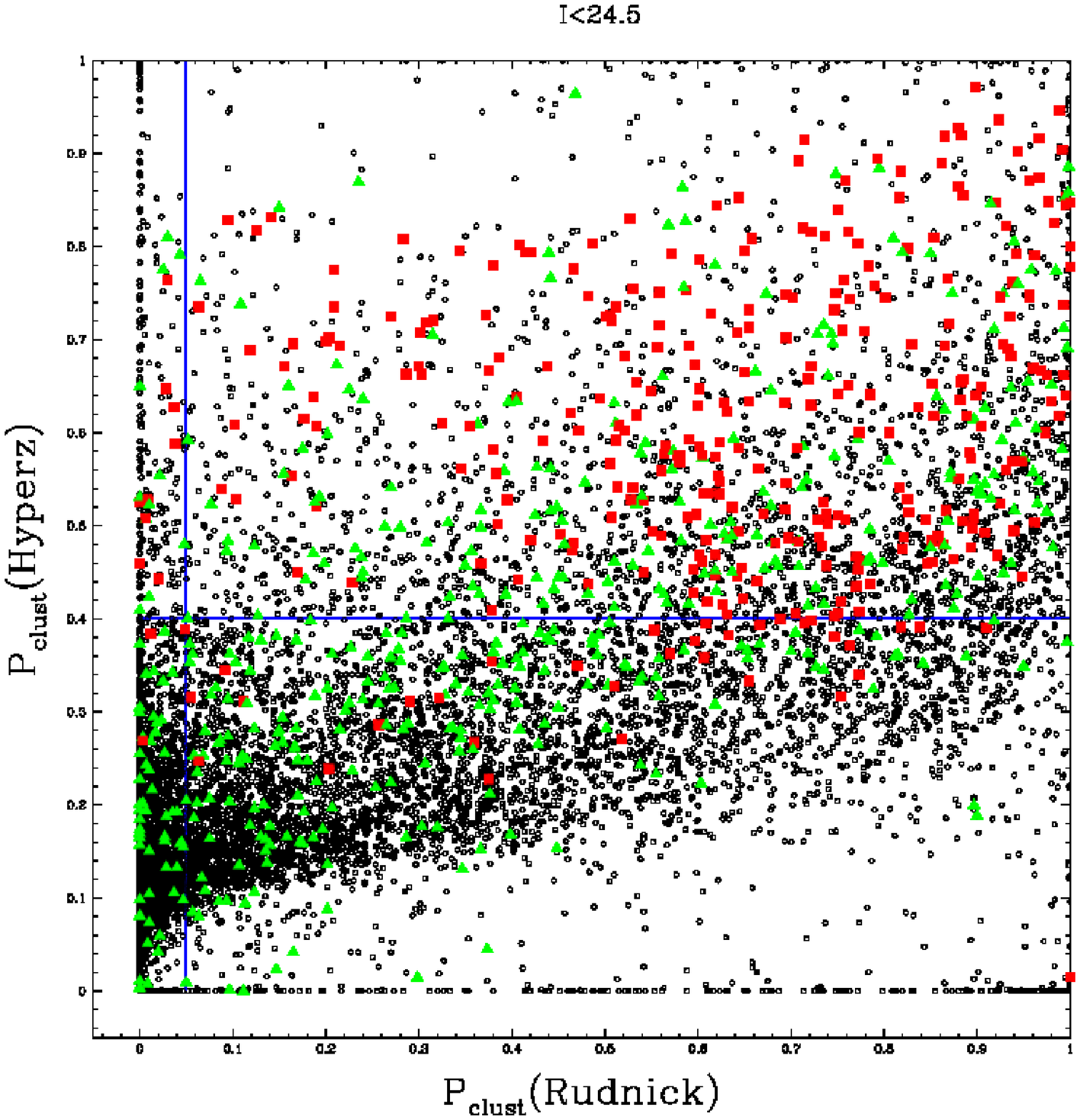,width=0.50\textwidth}
\caption {
Same as Figure~\ref{pint_comp_lowz} but for 
the clusters with VRIJK photometry.
}
\label{pint_comp_highz}
\end{figure*}

Since the performance is similar using {\it Hyperz}, we use both codes
jointly to provide the most efficient rejection. In
Figs.~\ref{pint_comp_lowz} and \ref{pint_comp_highz}, we plot
$P_{clust}(\mathrm{GR})$ versus $P_{clust}(\mathrm{Hyperz})$. While
there is a large scatter, there is a definite correlation between the
two probabilities, such that the majority of objects with low \pclust
for one code also have a low \pclust with the other code. A Spearman's
Rank Correlation test on the distribution of $P_{clust}(\mathrm{GR})$
vs. $P_{clust}(\mathrm{Hyperz})$ shows that there is higher than
99.9\% probability that these two variables are correlated. The same
result is found for the low and high-z samples, when using the full
magnitude-limited samples or other subsamples restricted to the
brightest galaxies. This implies that a joint rejection is feasible.
After extensive tests, we decided to reject galaxies if
\pclust$<$\pthresh for either code; in these tests, the \pthresh
values for each code were determined separately, such that the highest
rejection, independent of rest-frame color, was possible, while
retaining $>$90\% of the confirmed members.  The adopted thresholds
are summarized in Table~\ref{rejthreshtab} and the performance of
these thresholds is summarized in Table~\ref{rejfracspectab}.

%% New table 

\begin{table}
\caption{\label{rejthreshtab} Rejection thresholds}

\begin{tabular}{cccccc}
\hline \hline
Code  & Low-z & & & High-z & \\
      & $BVI$ & $BVIK$ & $BVIJK$ & $VRI$ & $VRIJK$ \\
\hline
 &  &  &  &  & \\
GR code & 0.350 & 0.475 & 0.300 & 0.200  & 0.050 \\
{\it Hyperz} & 0.150 & 0.425 & 0.425 & 0.050  & 0.400 \\
 &  &  &  &  & \\
\hline
\end{tabular}
\end{table}

%%% end new table 

%% New table 

\begin{table}
\caption{\label{rejfracspectab} Retained and Rejected fraction in
  Spectroscopic Sample}

\begin{tabular}{lccccc}
\hline \hline
      & Low-z & & & High-z & \\
      & $BVI$ & $BVIK$ & $BVIJK$ & $VRI$ & $VRIJK$ \\
\hline
 &  &  &  &  & \\
$f_{retain}^{memb}$   & 0.95 & 0.89 & 0.90 & 0.90 & 0.89 \\
 &  &  &  &  & \\
$f_{reject}^{nonmemb}$ & 0.53 & 0.73 & 0.88 & 0.44 & 0.53 \\
 &  &  &  &  & \\
\hline
\end{tabular}
\end{table}

%%% end new table 

Because \pz is broader for galaxies without NIR data, \pclust is also
systematically lower and the \pthresh, determined for galaxies with
NIR data, is no longer be applicable. To calibrate \pthresh for
galaxies without NIR data, we re-derived \zp for the entire
spectroscopic sample, excluding the NIR filters. We recalibrate
\pthresh and summarize the performance and adopted cuts in
Table~\ref{rejthreshtab} and \ref{rejfracspectab}. It is important to
note that the performance of the rejection is different in areas with
and without NIR data and the retained member population differs in the
two regions. For this reason, we limit all studies using the
photometric redshifts to those areas with NIR data.

We checked how the effectiveness of the adopted \pthresh varied across
the sample. Because of the limited numbers of spectroscopically
observed objects per cluster, this was not possible on a
cluster-by-cluster basis. Instead we split each of the high-z and
low-z samples into two subsamples each and examined how the accepted
and rejected fractions differed. The retained fractions of members
ranges from 87--98\% and the rejected fractions of non-members ranges
from 50--60\%.

%%%%%%%%%%%%%%%%%%%%%%%%%%%%%%%%%%%%%%%%%%%%%%%
\subsection{\pclust Threshold Performance in Magnitude Limited Samples}
\label{prob_maglim}

\begin{figure}
\psfig{file=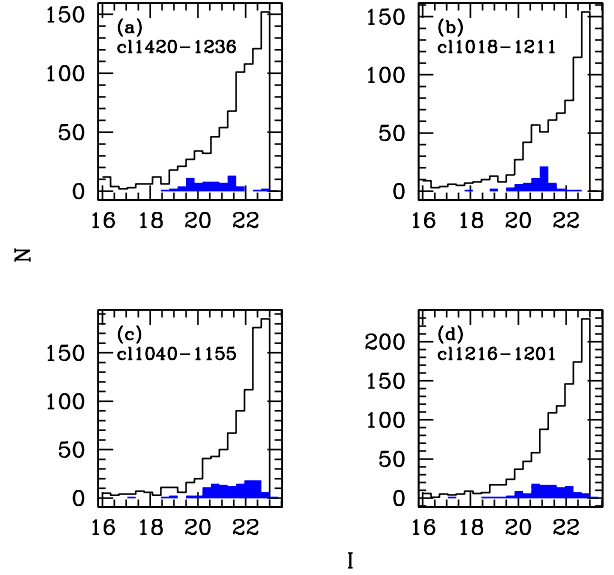,width=0.5\textwidth}
\caption {The apparent magnitude histograms for the spectroscopic
  sample (solid blue histogram) versus the total sample of the same
  limiting magnitude (open black). For both samples at low-z (a and b)
 and high-z (c and d), 
  two of the clusters with the most complete spectroscopic
  coverage are displayed. We note that the spectroscopic sample, even
  for clusters with the most extensive spectroscopy, is not complete
  at any magnitude limit and is biased towards brighter magnitudes.
  The magnitudes correspond to the $I$-band AUTO magnitudes from
  SExtractor.}
\label{maghist}
\end{figure}

We examine how applicable our adopted \pthresh, calibrated using the
spectroscopic subsample, is to the full magnitude-limited sample. In
Fig.~\ref{maghist}, we show the apparent magnitude distribution of the
spectroscopic sample and the total photometric sample with the same
magnitude limit, for two clusters in each redshift range with the
widest spectroscopic coverage. It is clear from these plots that we
are not spectroscopically complete at any magnitude and that the
spectroscopic sample, as expected, is biased towards brighter
magnitudes.

\begin{figure*}[!h]
\psfig{file=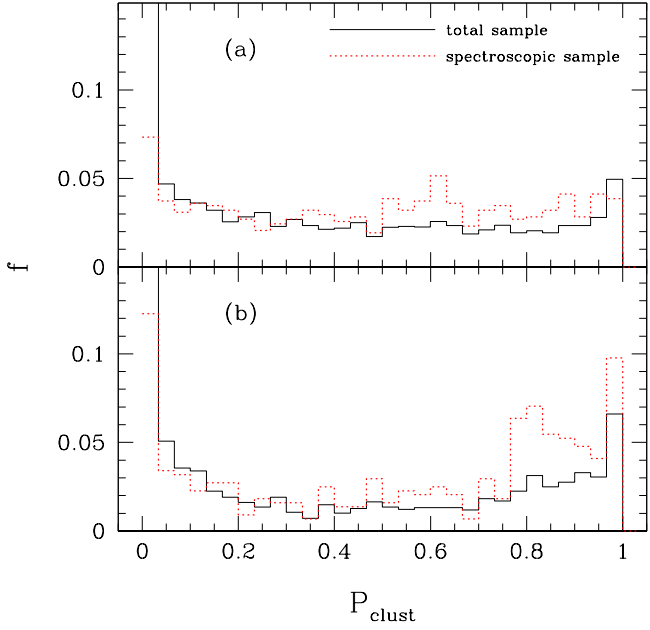,width=0.50\textwidth}
\caption {The histogram of \pclust values for the photometric sample
  (solid line) versus that of the spectroscopic sample (dotted line).
  Panel (a) is for the high-z sample at $I<22.5$ and panel (b) is for the
  low-z sample at $I<22$. In each case, the histograms represent the
  fraction of the total number of objects over all \pclust.}
\label{pint_hist}
\end{figure*}

The spectroscopic target lists were not only constructed with a
magnitude limit in mind, but also with an eye towards reducing the
number of galaxies that had low probabilities of being at the cluster
redshift, in addition to including some galaxies that were not
formally present about \zclust. We examine in general how this
preselection causes the \pclust distributions of the spectroscopic
sample to differ from those in a magnitude-limited photometric
sample. In Fig.~\ref{pint_hist}, we compare the histogram of the
\pclust values for the spectroscopic samples with those for the entire
sample down to the same magnitude limits. The spectroscopic
preselection manifests itself as an excess of high \pclust values and
a deficit of low \pclust for the spectroscopic sample with respect to
the photometric sample. 
Within the precision of our numerical routine, a KS-test gives 0\%
probability that these two distributions are drawn from the same parent
distribution.
This inherent bias implies that a certain
\pthresh removes a higher fraction of galaxies in the photometric
sample than was the case in the spectroscopic subsample. 

\begin{figure}
\psfig{file=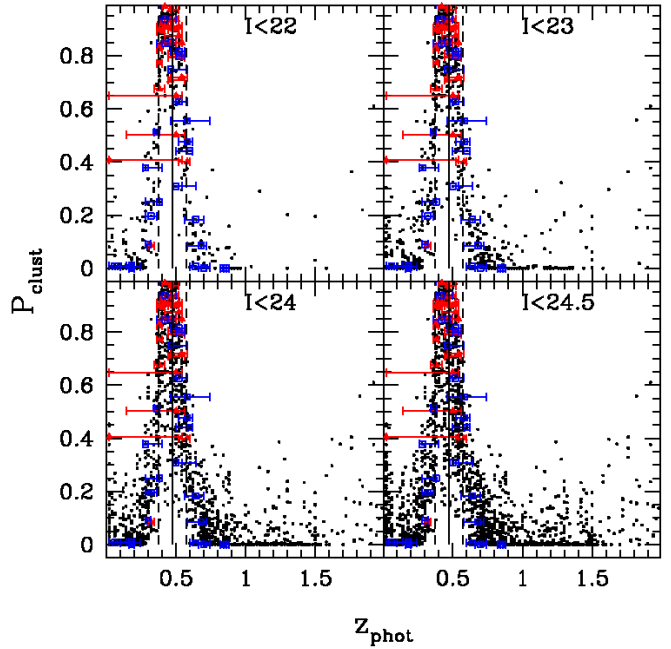,width=0.5\textwidth}
\caption {A plot of \pclust vs. \zp for Cl1018-1211 that compares the
  spectroscopic sample with the magnitude-limited sample. Each tile
  corresponds to a different magnitude limit for the photometric
  sample. The full photometric sample is indicated by black dots.
The spectroscopically confirmed cluster members are
  indicated by solid red triangles and the non-members by open blue
  squares. The solid vertical line indicates \zp=\zclust and the
  dashed lines indicate the intervals $\pm0.1$ in redshift. We note
  that, at fainter magnitudes, the galaxies with low \pclust values do
  not lie at $z_{phot}\sim z_{clust}$ in large numbers, but rather are
  at different redshifts.}
\label{probzcomp_cl1018}
\end{figure}

\begin{figure}
\psfig{file=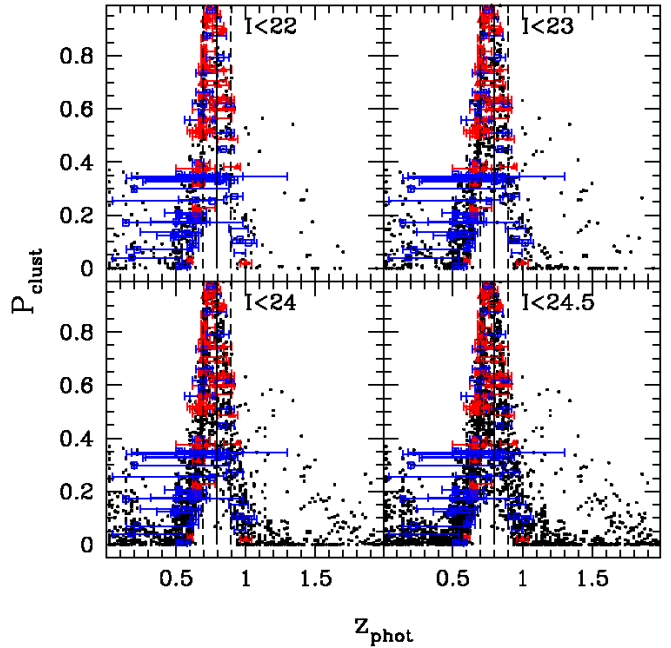,width=0.5\textwidth}
\caption {Same as Figure~\ref{probzcomp_cl1018} except for
  Cl1216-1201.}
\label{probzcomp_cl1216}
\end{figure}

\begin{figure*}[!h]
\psfig{file=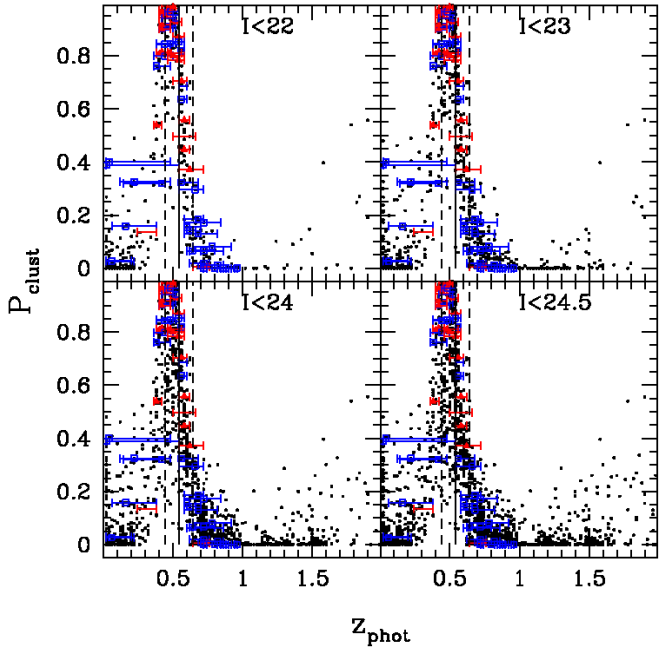,width=0.50\textwidth}
\caption {Same as Figure~\ref{probzcomp_cl1018} except for
  Cl1420-1236.}
\label{probzcomp_cl1420}
\end{figure*}

\begin{figure*}[!h]
\psfig{file=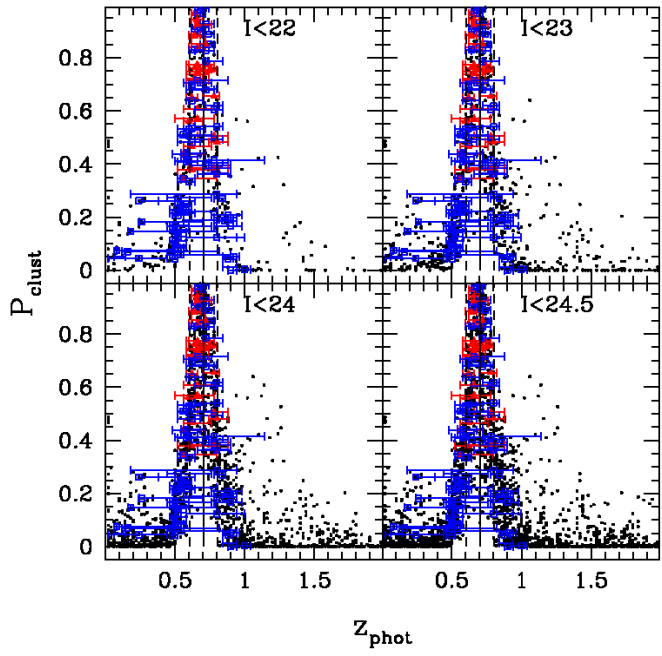,width=0.50\textwidth}
\caption {Same as Figure~\ref{probzcomp_cl1018} except for
  Cl1040-1155.}
\label{probzcomp_cl1040}
\end{figure*}

We examine how the spectroscopically calibrated \pclust rejection
operates when applied to a magnitude-limited sample. We illustrate these
results using the GR code, but the conclusions would be equivalent using
{\it Hyperz}. In Figs.~\ref{probzcomp_cl1018},
\ref{probzcomp_cl1420}, \ref{probzcomp_cl1216}, and
\ref{probzcomp_cl1040}, we plot the \pclust versus \zp for two high-z
and two low-z clusters. Each figure has panels that show how this
distribution changes with apparent magnitude. As we move to fainter
magnitude limits, many galaxies appear at all \pclust values. Those at
high \pclust do indeed fall close to \zclust, as predicted by the
spectroscopic studies. Encouragingly, the galaxies with low \pclust
values fall systematically away from \zclust.  In fact, even at the
faintest magnitudes, there are very few galaxies within
\zclust$\pm0.1$ that have \pclust$<0.2$. We recall that this must not
be the case: galaxies with \zp$\approx$\zclust but broad \pz
distributions will have low probabilities of being at the cluster
redshift, even though their best value lies around \zclust. This
self-consistency implies that \pclust is, in fact, providing us with a
real indication of whether these faint objects are at the cluster
redshift. This effect would be difficult to reproduce by systematic
errors in the photometric redshifts because it appears for all
clusters, regardless of their redshift range or presence of NIR data.
Nonetheless, there are some faint galaxies with $z_{phot}\sim
z_{clust}$ that may be rejected because of a broad P(z).  Also, the
tests presented in Sect.~\ref{zzfaint} demonstrate that the
photometric redshift accuracy is expected to be lower for fainter
galaxies, implying that there will be galaxies who are truly at
$z_{cluster}$ but are scattered away from the cluster redshift. In
Rudnick et al.\ 2009, we discuss how these effects may differ
for red and blue galaxies and we present the implications for the
study of the cluster galaxy luminosity function.

The total fraction of galaxies rejected for each cluster as a function
of $I$ magnitude are presented in the columns 3-6 of
Table~\ref{tbl:rejfrac_maglim.tab}. When our rejection criteria is
applied, we reject 55-82\% of the galaxies at $I<22$ and 75-93\% at
$I<24.5$.

%% Table rejections 
\begin{table}[h]
\caption{\label{tbl:rejfrac_maglim.tab}
Total fraction of galaxies rejected as cluster members as a function of
the $I$ magnitude (columns 3 to 6 for $I\ge$ 22, 23, 24 and 24.5
respectively). }
\begin{tabular}{lccccc}
\hline \hline
Cluster & $z_{cl}$ & $f_{22}$ & $f_{23}$ & $f_{24}$ & $f_{24.5}$ \\
\hline
Cl1018.5-1211 & 0.4734 & 0.68 & 0.77 & 0.85 & 0.88 \\
Cl1037.5-1243a & 0.4252 & 0.50 & 0.58 & 0.71 & 0.77 \\
Cl1040.4-1156 & 0.7043 & 0.68 & 0.73 & 0.83 & 0.86 \\
Cl1054.2-1146 & 0.6972 & 0.56 & 0.67 & 0.78 & 0.83 \\
Cl1054.4-1245 & 0.7498 & 0.58 & 0.60 & 0.69 & 0.75 \\
Cl1059.1-1253 & 0.4564 & 0.68 & 0.76 & 0.84 & 0.87 \\
Cl1103.4-1245b & 0.7031 & 0.72 & 0.77 & 0.83 & 0.86 \\
Cl1119.2-1129 & 0.5500 & 0.69 & 0.75 & 0.81 & 0.83 \\
Cl1122.5-1136 & 0.6397 & 0.61 & 0.66 & 0.79 & 0.84 \\
Cl1138.1-1133 & 0.4796 & 0.53 & 0.62 & 0.74 & 0.79 \\
Cl1202.4-1224 & 0.4240 & 0.80 & 0.86 & 0.91 & 0.93 \\
Cl1216.4-1201 & 0.7943 & 0.62 & 0.66 & 0.74 & 0.78 \\
Cl1227.5-1138 & 0.6357 & 0.58 & 0.64 & 0.73 & 0.77 \\
Cl1232.3-1250 & 0.5414 & 0.67 & 0.73 & 0.82 & 0.85 \\
Cl1238.3-1144 & 0.4602 & 0.65 & 0.73 & 0.81 & 0.83 \\
Cl1301.4-1139 & 0.4828 & 0.71 & 0.79 & 0.87 & 0.90 \\
Cl1353.0-1137 & 0.5882 & 0.77 & 0.84 & 0.89 & 0.91 \\
Cl1354.1-1231 & 0.7620 & 0.64 & 0.66 & 0.75 & 0.78 \\
Cl1411.0-1148 & 0.5195 & 0.81 & 0.84 & 0.90 & 0.92 \\
Cl1420.2-1236 & 0.4962 & 0.82 & 0.87 & 0.91 & 0.93 \\
\hline
\end{tabular}

Notes -- 
Only galaxies with NIR data were used in this calculation,
except for Cl1119-1129 and Cl1238-1144. For these two
clusters, all galaxies were used because no NIR data exists.
\end{table}

%%%%%%%%%%%%%%%%%%%%%%%%%%%%%%%%%%%%%%%%%%%%%%%
%%%%%%%%%%%%%%%%%%%%%%%%%%%%%%%%%%%%%%%%%%%%%%%
\section{\label{conclusions} Discussion and Conclusions}

We have used two independent codes to compute photometric redshifts:
{\it Hyperz} and GR code. In general, the two codes yield rather
similar results, either on a cluster-by-cluster basis or as a function
of the filter set and spectral type, of typically $\sigma(\Delta
z/(1+z))\sim 0.05$ to 0.06. {\it Hyperz} results are found to be
slightly more accurate than GR's ones in general, by $\ltapprox$20\%
in $\sigma(\Delta z/(1+z))$. The quality achieved by both codes is
consistent with the expectations derived from ``ideal'' simulations.
An interesting trend is that the quality of both codes is highly
correlated, in the sense that the highest and lowest quality results,
in terms of $\sigma(\Delta z/(1+z))$, and systematics are found for
the same clusters. This trend cannot be due to the use of an
incomplete or imperfect template set, as suggested by other authors
(Ilbert et al.\ 2006), because in such a case, we should expect the
same systematic behavior in all fields, given a filter set, as
discussed in \S~\ref{simus}. In contrast, different systematics are
observed in the different fields, which are found to be almost equal
for the two independent \zphot codes. This behavior suggests that the
origin of the systematic errors is more likely to be associated with
small residuals in the input photometry rather than the \zphot
templates and codes. Indeed, small zero-point shifts of
$\ltapprox$0.05 magnitudes cannot be excluded, in particular for the
near-IR data.
   
Photometric redshifts are found to be particularly useful in the
identification and study of galaxy clusters in large surveys. The
determination of cluster redshifts in the EDisCS fields using a simple
algorithm based on \zphot is highly accurate. Indeed, the differences
between photometric and spectroscopic values are found to be small,
typically ranging between $\delta z \sim$0.03-0.04 in the high-z
sample and $\delta z \sim$0.05 in the low-z sample. This is at least a
factor $\sim(1+z)$ more accurate than the determination of \zphot for
individual galaxies. The accuracy is more sensitive to the filter set
used rather than the redshift of the cluster. The systematic lower
quality results for the low-z sample was somewhat expected from the
simulations presented in \S~\ref{simus}.
Tomography based on \zphot could be used in searches for clusters
along the line-of-sight, using redshift steps optimized to be close in
value to the typical difference between photometric and spectroscopic
z$_{cluster}$ to maximize the contrast between members and non-member
galaxies (in this case, $\Delta z \sim 0.05$).

The cluster membership criterion presented in Sect.~\ref{member} has
been used to extend the spectroscopic studies of cluster galaxies to
fainter limits in magnitude (e.g. De Lucia et al.\ 2004, White et al.\
2005, Clowe et al.\ 2006, Poggianti et al.\ 2006, De Lucia et al.\
2007, Desai et al.\ 2007, Rudnick et al.\ 2009).

In conclusion, photometric redshifts are useful tools for studying
galaxy clusters. They enable efficient and complete pre-selection of
cluster members for spectroscopy, allow accurate determinations of the
cluster redshifts based on photometry alone, provide a means of
determining cluster membership, especially for bright sources, and can
be used to search for galaxy clusters.

%%%%%%%%%%%%%%%%%%%%%%%%%%%%%%%%%%%%%%%%%%%%%%%
\acknowledgements

Part of this work was supported by the French {\it Centre National de
  la Recherche Scientifique}, by the French {\it Programme National de
  Cosmologie} (PNC) and the {\it Programme National Galaxies} (PNG).
The Dark Cosmology Centre is funded by the Danish National Research
Foundation. 

%%%%%%%%%%%%%%%%%%%%%%%%%%%%%%%%%%%%%%%%%%%%%%%
%%%%%%%%%%%%%%%%%%%%%%%%%%%%%%%%%%%%%%%%%%%%%%%

%%%%%%%%%%%%%%%%%%%%%% ONLINE MATERIAL %%%%%%%%%%%%%%%%%%%%%%%%%%%%%%%%
%%%%%%%%%%%%%%%%%%%%%%%%%%%%%%%%%%%%%%%%%%%%%%%%%%%%%%%%%%%%%%%%%%%%%%%

\Online

% appendix
%\appendix

%%% Online tables

%%%%%%%%%%%%%%%% Table simus GR code %%%%%%%%%%%%%%%% 

\begin{table*}
\begin{center}
\caption{\label{GR_zzsimulated}\zphot accuracy derived from simulations for
  galaxies in the spectroscopic sample, based on the GR code.}
\begin{tabular}{clccccccc}
\hline \hline
Clusters & Redshift interval & Galaxy type &
$\left<\Delta_z \right>$ & $\sigma_z$ &
$\sigma_{z,MAD}$ & $\sigma(\Delta z/(1+z))$ & $l\%$ & $g\%$  \\
\hline 
Low-z & $0.3 \le$  \zphot$ \le 1$ &  all & 
0.008   &  0.150 & 0.104  &   0.063 & 0.9 &    7.7 \\
High-z & $0.3 \le$  \zphot$ \le 1$ & all &
-0.072 &    0.164 & 0.167  &   0.100 & 10.6  &   20.3 \\
High-z & $0.45 \le$  \zphot$ \le 1$ & all &  
-0.092 &   0.158 & 0.151 &  0.088 & 6.5 &    21.5 \\
\hline
Low-z & $0.3 \le$  \zphot$ \le 1$ &  E/S0 & 
-0.140  &   0.129 & 0.184  &   0.110 & 4.2 &    6.3 \\
      &                             & Sbc & 
-0.077  &   0.117  & 0.136   &   0.082 & 0.8 &     9.9 \\
      &                             & Scd & 
0.013  &    0.076 & 0.070  &    0.041 &  0.0  &    4.3 \\
      &                             & Im & 
0.024 &    0.077 & 0.075   &  0.045 & 0.0 &    5.9 \\
      &                             & SB &
0.098  &   0.117 & 0.121  &   0.076 & 0.1 &   9.8 \\
\hline
High-z & $0.3 \le$  \zphot$ \le 1$ & E/S0 & 
-0.227  &   0.153 & 0.370  &   0.226 & 12.2 &    31.6 \\
      &                             & Sbc & 
-0.101  &    0.161  & 0.232  &    0.134 & 17.3   &   28.0 \\
      &                             &  Scd & 
-0.025  &   0.093 & 0.086  &   0.052 & 2.7  &   16.2  \\
      &                             & Im & 
0.024   &  0.126 & 0.123  &   0.074 & 5.9  &   6.7 \\
      &                             & SB & 
-0.045  &   0.063 & 0.077  &   0.047 & 2.3  &   13.5 \\
\hline 
\end{tabular}
\end{center}
Notes --
 The accuracy reached for
  the different spectral types of galaxies is also presented. The
  information given is the same as in Table~\ref{zzsum} and
  Table~\ref{zzsimulated}, for the current spectroscopic sample. 
\end{table*}

%%%%%%%%%%%%%%%% end Table simus GR %%%%%%%%%%%%%%%% 

%%%%%%%%%%%%%%%% Table simulations faint sample (GR) %%%%%%%%%%%%%%%% 
\begin{table*}
\begin{center}
\caption{\label{GR_tab_zzfaint}\zphot accuracy expected for the faintest galaxies in the
  EDisCS sample, based on the GR code. } 

\begin{tabular}{clccccccc}
\hline \hline
Clusters & Redshift interval & Galaxy type &
$\left<\Delta_z \right>$ & $\sigma_z$ &
$\sigma_{z,MAD}$ & $\sigma(\Delta z/(1+z))$ & $l\%$ & $g\%$  \\
\hline 
Low-z & $0.3 \le$  \zphot$ \le 1$ &  all & 
0.034  &   0.136 & 0.124 &    0.082 & 3.1 &    0.0 \\
High-z & $0.3 \le$  \zphot$ \le 1$ & all &
0.003  &    0.131&  0.134  &    0.081&  3.4&   2.9\\
High-z & $0.45 \le$  \zphot$ \le 1$ & all &  
0.016 &     0.124 & 0.130 &     0.078 & 3.4 & 3.0 \\
\hline
Low-z & $0.3 \le$  \zphot$ \le 1$ &  E/S0 & 
0.015 & 0.123 & 0.118 & 0.078 & 1.8 & 0.0  \\
      &                             & Sbc & 
0.043 & 0.128 & 0.124 & 0.081 & 0.7 & 0.0 \\
      &                             & Scd & 
0.036 & 0.145 & 0.112 & 0.075 & 6.4 & 0.0 \\
      &                             & Im & 
0.093 & 0.179 & 0.153 & 0.103 & 8.9 & 0.0  \\
      &                             & SB &
0.041 & 0.145 & 0.129 & 0.083 & 6.7 & 0.0 \\
\hline
High-z & $0.3 \le$  \zphot$ \le 1$ & E/S0 & 
0.014  &   0.129 &0.149&     0.088 & 1.2& 1.6 \\
      &                             & Sbc & 
-0.010 &   0.125 &  0.128 &     0.080 & 0.9 & 1.3  \\
      &                             &  Scd & 
-0.008 &    0.134 &0.114 &    0.067 & 4.0 & 4.7 \\
      &                             & Im & 
0.012  &   0.124 & 0.122 &    0.073 & 5.4 & 0.0 \\
      &                             & SB & 
0.011 &    0.140 & 0.134  &   0.092 & 13.6 & 11.7 \\
\hline 
\end{tabular}
\end{center}
Notes -- 
The accuracy reached for the different spectral types of
  galaxies is also presented.  The information given is the same as in
  Table~\ref{zzsum} and Table~\ref{tab_zzfaint}, for the current spectroscopic
  sample.  
\end{table*}
%%%%%%%%%%%%%%%% end Table faint GR %%%%%%%%%%%%%%%% 

%%% Online Figures

%%%%%%%%%%%%%%%%%%%%%%%%%%%%%%% old Fig. 5: Low-z sample

%%%%%%%%%%%%%%%%%%%%%%%%%%%%%%% old Fig. 7: High-z sample

%%%%%%%%%%%%%%%%%%%%%%%%%%%%%%%%%%%%%%%%%%%%%%%%%%%%%%%%%%%%%%%%%%%%%%%

%%%%%%%%%%%%%%%%%%%%%%%%%%%%%%%%%%%%%%%%%%%%%%%%%%%%%%%%%%%%%%%%%%%%%%%
% old Fig. 18

%%%%%%%%%%%%%%%%%%%%%%%%%%%%%%%%%%%%%%%%%%%%%%%%%%%%%%%%%%%%%%%%%%%%%%%
% old Fig. 20 and 22

%%%%%%%%%%%%%%%%%%%%%%%%%%%%%%%%%%%%%%%%%%%%%%%%%%%%%%%%%%%%%%%%%%%%%%%

%% end online material

%%%%%%%%%%%%%%%%%%%%%%%%%%%%%%%%%%%%%%%%%%%%%%%%%%%%%%%%%%%%%%%%%%%%%%%
%%%%%%%%%%%%%%%%%%%%%%%%%%%%%%%%%%%%%%%%%%%%%%%%%%%%%%%%%%%%%%%%%%%%%%%

\end{document}

%% file: 10644_t1.tex
{\scriptsize

\begin{tabular}{clcccc|cccccc|cccccc}
\hline \hline
 &  &  &  &  &  & Hyperz &  &  &  &  & & GR &  &  &  &  &  \\
\hline 

Cluster (Low-z) & Filters & z$_{cluster}$ & E(B-V) & N$_{total}$ & N(1) & 
$\left<\Delta_z \right>$ & $\sigma_z$ & 
$\sigma_{z,MAD}$ & $\sigma(\Delta z/(1+z))$ & $l\%$ & $g\%$ & 
$\left<\Delta_z \right>$ & $\sigma_z$ &  
$\sigma_{z,MAD}$ & $\sigma(\Delta z/(1+z))$ & $l\%$ & $g\%$ \\
(1) & (2)  & (3) & (4) & (5) & (6) & (7)& (8) & (9) &(10)  & (11) & (12)& (13) & (14) & (15) & (16) & (17) & (18)\\
  &  &  &  &  &  & &  &  &  &  & &  &  &  &  &  & \\

\hline 
cl1018-1211  & BVIK$_s$  & 0.473  & 0.0773& 63 & 56 & 
0.037  & 0.060 & 0.083 & 0.054 & 0.0 & 0.0 
& 0.023 & 0.085 & 0.095& 0.064 & 1.8 & 0.0 \\
cl1059-1253  & BVIK$_s$  & 0.456  & 0.0332& 78 & 67 & 
0.018 &  0.064 & 0.069 & 0.045 &  0.0 & 2.9
& 0.015 & 0.082 & 0.069 & 0.047 & 0.0 & 1.5 \\
{\it cl1119-1129 } & {\it BVI   }    & {\it 0.550 } & {\it 0.0332} & {\it 57 }& {\it 45} & 
{\it 0.061 }& {\it 0.117 }&{\it  0.133 }& {\it 0.090 }& {\it 11.1 }& {\it 0.0}
& {\it 0.032 }& {\it 0.089 }& {\it 0.103 }& {\it 0.066 }& {\it 11.1 }& {\it 0.0 }\\
cl1202-1224  & BVIK$_s$  & 0.424  & 0.0583& 59 & 46 & 
-0.007  & 0.111 & 0.113 & 0.079 & 0.0 & 0.0
& 0.041 & 0.091 & 0.090 & 0.062 & 0.0 & 0.0 \\
cl1232-1250  & BVIJK$_s$ & 0.541  & 0.0596& 93 & 77 & 
-0.013 & 0.083 & 0.060 & 0.039 & 1.3 & 2.7 
& 0.022 & 0.087 & 0.097 & 0.066 & 0.0 & 2.5 \\
{\it cl1238-1144}  & {\it BVI }      & {\it 0.460 } & {\it 0.0437}& {\it 11} & {\it 8 }& 
{\it -0.011 }& {\it 0.072 }& {\it 0.103}& {\it 0.072 }& {\it 0.0} & {\it 0.0}
& {\it 0.008 }& {\it 0.054 }& {\it 0.065 }& {\it 0.043} & {\it 12.5 }& {\it 0.0  } \\
cl1301-1139  & BVIK$_s$  & 0.482  & 0.0487& 78 & 69 & 
-0.012 & 0.062 & 0.058& 0.039 & 0.0 & 0.0
& -0.004 & 0.078 & 0.081& 0.054& 0.0 & 1.4 \\
cl1353-1137  & BVIK$_s$  & 0.588  & 0.0487& 62 & 55 & 
0.064 & 0.142 & 0.104& 0.066 & 0.0 & 0.0
& 0.045 & 0.103 & 0.110& 0.071 & 0.0 & 0.0 \\
cl1411-1148  & BVIK$_s$  & 0.520  & 0.0654& 68 & 63 & 
0.017 & 0.112 & 0.086& 0.057 & 0.0 & 0.0
& 0.048 & 0.128 & 0.089& 0.058& 0.0 & 0.0 \\
cl1420-1236  & BVIK$_s$  & 0.496  & 0.0821& 68 & 58 & 
0.007 &  0.088 & 0.069&0.046 & 0.0 & 0.0 
& -0.001& 0.093 & 0.091 & 0.060& 0.0 & 0.0 \\
\hline
Average &  &  &   &    &    & 
0.014 & 0.090 & 0.080 & 0.053
&0.1 & 0.7 &
0.024& 0.093& 0.090& 0.060
&0.2 & 0.7 \\ 
rms &  &  &   &    &    & 
0.025 & 0.027 &0.019 &0.013
&0.4 & 1.2 &
0.019& 0.015 & 0.011 & 0.007
&0.6 & 1.0 \\ 
%%      &  &  &   &    &    & 
%% $\pm$0.025 & $\pm$0.027 & $\pm$0.019 & $\pm$0.013
%% &$\pm$0.4 & $\pm$1.2 &
%% $\pm$0.019& $\pm$0.015 & $\pm$0.011 & $\pm$0.007
%% &$\pm$0.6 & $\pm$1.0 \\ 
%% 
%%   &  &  &  &  &  & &  &  &  &  & &  &  &  &  &  & \\
%%  
\hline \hline
Cluster (high-z) & Filters & z$_{cluster}$ & E(B-V) & N$_{total}$ & N(1) & $\left<\Delta_z \right>$ &
$\sigma_z$ &  
$\sigma_{z,MAD}$ & $\sigma(\Delta z/(1+z))$ & $l\%$ & $g\%$ & 
$\left<\Delta_z \right>$ & $\sigma_z$ &  
$\sigma_{z,MAD}$ & $\sigma(\Delta z/(1+z))$ &$l\%$ & $g\%$ \\
(1) & (2)  & (3) & (4) & (5) & (6) & (7)& (8) & (9) &(10)  & (11) & (12)& (13) & (14) & (15) & (16) & (17) & (18)\\
  &  &  &  &  &  & &  &  &  &  & &  &  &  &  &  & \\
\hline 
cl1037-1243a  & VRIJK$_s$  & 0.425  & 0.0433& 111  & 104 & 
-0.016 & 0.114 & 0.107 & 0.075& 0.0 & 1.0
& -0.010 & 0.092 &0.098 & 0.065 & 1.9 & 1.9 \\
cl1040-1155  & VRIJK$_s$  & 0.704  & 0.0505& 116 & 104 & 
0.028 & 0.068 & 0.080& 0.046 & 0.0 & 1.9
& 0.021 & 0.072 & 0.076& 0.046 & 0.0 & 1.9 \\
cl1054-1146  & VRIJK$_s$  & 0.697  & 0.0374& 102 & 94 & 
0.025  & 0.090  & 0.090& 0.054 & 0.0 & 3.3
& 0.020 & 0.127 & 0.114& 0.068 & 2.1 & 4.3 \\
cl1054-1245  & VRIJK$_s$  & 0.750  & 0.0378& 98 & 77 & 
-0.001 & 0.087 & 0.094  & 0.055 &0.0 & 5.8
& 0.063& 0.071 & 0.111 & 0.064 & 0.0 & 7.0 \\
cl1103-1245b  & VRIJK$_s$  & 0.703  & 0.0481& 98 & 86 & 
0.012 &  0.088 & 0.085& 0.052 & 2.3 & 4.1
& -0.004 & 0.121 & 0.112 & 0.069& 7.0 & 8.4 \\
{\it cl1122-1136}  & {\it VRIJK$_s$ } & {\it 0.640}  & {\it 0.0404}& {\it 12 }&
          {\it 7}  & 
{\it -0.012 }& {\it 0.105 }&{\it 0.180}&{\it 0.107 }& {\it 14.3 }& {\it 14.3}
& {\it -0.005 }& {\it 0.115 }& {\it 0.061 }& {\it 0.042}& {\it 0.0} & {\it 0.0 } \\
cl1138-1133  & VRIJK$_s$  & 0.479  & 0.0274& 110 & 94 & 
-0.002 &  0.088 & 0.088& 0.056 & 0.0 & 1.1
& -0.035 & 0.080 & 0.080 & 0.054& 0.0 & 2.0 \\
cl1216-1201  & VRIJK$_s$  & 0.794  & 0.0449& 116 & 96 & 
0.034 & 0.079 & 0.096 & 0.053 & 1.0 & 7.8
& 0.040 & 0.096 & 0.136& 0.077& 0.0 & 7.6 \\
cl1227-1138  & VRIJK$_s$  & 0.635  & 0.0468& 110 & 100 &  
-0.030 & 0.115 & 0.087& 0.054  & 1.0 & 2.0
& -0.004 & 0.101& 0.075& 0.047 & 0.0 & 3.0 \\
cl1354-1230  & VRIJK$_s$  & 0.762  & 0.0793& 104 & 92 & 
-0.018 & 0.108 & 0.090& 0.053 & 3.3 & 2.2
& 0.010 & 0.107 & 0.086 & 0.052 & 1.1& 1.1 \\
\hline 
Average  &  &  &   &    &    & 
0.003 & 0.093 & 0.091 & 0.055 & 0.8 & 3.2 &
0.011 & 0.096 & 0.099 & 0.060 & 1.3 & 4.1 \\
rms &  &  &   &    &    & 
0.021 & 0.015 & 0.007 & 0.007 &1.1 & 2.2 &
0.027 & 0.019 & 0.020 & 0.010 & 2.1 & 2.6 \\
%%        &  &  &   &    &    & 
%%  $\pm$0.021 & $\pm$0.015 & $\pm$0.007 & $\pm$0.007 & $\pm$1.1 &$\pm$2.2&
%%  $\pm$0.027 & $\pm$0.019 & $\pm$0.020 &$\pm$0.010 &  $\pm$2.1 & $\pm$2.6 \\
%%    &  &  &  &  &  & &  &  &  &  & &  &  &  &  &  & \\
\hline \hline
\end{tabular}

}

%%% Local Variables: 
%%% mode: latex
%%% TeX-master: t
%%% End: 

%% file: 10644_t2.tex
{\scriptsize

\begin{tabular}{clccc|cccccc|cccccc}
\hline \hline
  &  &  &  &  & Hyperz &  &  &  &  & & GR &  &  &  &  &  \\
\hline

Clusters & Redshift interval & \zspec quality & Galaxy type & N &
$\left<\Delta_z \right>$ & $\sigma_z$ &
$\sigma_{z,MAD}$ & $\sigma(\Delta z/(1+z))$ & $l\%$ & $g\%$ &
$\left<\Delta_z \right>$ & $\sigma_z$ &
$\sigma_{z,MAD}$ & $\sigma(\Delta z/(1+z))$ & $l\%$ & $g\%$ \\
(1) & (2)  & (3) & (4) & (5) & (6) & (7)& (8) & (9) &(10)  & (11) & (12)& (13) & (14) & (15) & (16) & (17) \\
%  &  &  &   & &  &  &  &  & &  &  &  &  &  &  & \\
\hline 
Low-z & $0.3 \le$  \zphot$ \le 1$ & 1 & all & 544 & 
0.017 & 0.103 & 0.078 & 0.052 & 1.1 & 0.7 &
0.030 & 0.107 & 0.088 & 0.058 & 1.3 & 0.8 \\
      &                           & 1 $+$ 2 & all & 564 & 
0.016 & 0.106 & 0.079 & 0.053 & 1.1 & 0.7 &
0.029 & 0.110 & 0.089 & 0.058 & 1.2 & 0.7 \\
\hline
High-z & $0.3 \le$  \zphot$ \le 1$ & 1 & all & 
{\bf 854 }&
0.003 & 0.099 & 0.090 & 0.054 & 0.9 & 3.3 &
0.012 & 0.100 & 0.094 & 0.058 & 1.3 & 4.0 \\
      &                           & 1 $+$ 2 & all &  
{\bf 885 } & 
0.004 & 0.100 & 0.090 & 0.055 & 1.1 & 3.3 &
0.010 & 0.104 & 0.095 & 0.058 & 1.4 & 3.9 \\
\hline
High-z & $0.45 \le$  \zphot$ \le 1$ & 1 & all &  736 &
0.013 & 0.087 & 0.087 & 0.051 & 0.7 & 3.6 &
0.024 & 0.086 & 0.088 & 0.053 & 0.4 & 4.1 \\
      &                             & 1 $+$ 2 & all &  754 & 
0.014 & 0.088& 0.087 & 0.052 & 0.9 & 3.5 &
0.023 & 0.089& 0.089 & 0.053 & 0.5 & 4.0 \\
\hline
Low-z & $0.3 \le$  \zphot$ \le 1$ & 
{\bf 1 $+$ 2 $+$ 3}
& E/S0 & 232 &
0.006 & 0.071& 0.069 & 0.045 & 0.0 & 0.0 &
0.036 & 0.085& 0.091 & 0.061 &  0.0 & 0.0 \\
      &                             & 
{\bf 1 $+$ 2 $+$ 3}
& Sbc & 167 &
0.027 & 0.128& 0.104 & 0.069 & 1.2 & 1.3 &
0.036 & 0.104& 0.092 & 0.066 & 0.6 & 1.3 \\
      &                             & 
{\bf 1 $+$ 2 $+$ 3}
& Scd & 113 &
0.027 & 0.103& 0.075 & 0.047& 2.7 & 0.0 &
-0.004 & 0.102& 0.077& 0.047& 4.4 & 0.9 \\
      &                             & 
{\bf 1 $+$ 2 $+$ 3}
& Im & 46 &
-0.007 & 0.104& 0.087 &0.056& 0.0 & 4.0 &
0.051 &  0.166& 0.102 &0.062& 0.0 & 0.0 \\
      &                             & 
{\bf 1 $+$ 2 $+$ 3}
& SB & 18 &
0.018 & 0.153& 0.159 & 0.094& 11.1 & 5.0 &
-0.019& 0.204& 0.167 & 0.105& 5.6 & 5.2 \\
\hline
High-z & $0.3 \le$  \zphot$ \le 1$ & 
{\bf 1 $+$ 2 $+$ 3}
& E/S0 & 263 &
0.040  & 0.070& 0.088 & 0.052& 0.8 & 0.8 &
0.025  & 0.114& 0.126 & 0.077& 0.8 & 1.9 \\
      &                             & 
{\bf 1 $+$ 2 $+$ 3}
& Sbc & 256 &
-0.009 & 0.103& 0.089 & 0.055& 1.6 & 1.6 &
0.005 &  0.108& 0.110 & 0.068& 1.6 & 2.7 \\
      &                             & 
{\bf 1 $+$ 2 $+$ 3}
& Scd & 181 &
0.017 &  0.094& 0.094 & 0.056& 1.1 & 1.7 &
-0.010&  0.079& 0.063 & 0.038& 2.8 & 7.8 \\
      &                             & 
{\bf 1 $+$ 2 $+$ 3}
& Im & 112 &
-0.012 & 0.079& 0.075 & 0.047& 0.0 & 0.9 &
0.024  & 0.066& 0.075 & 0.044&0.0 & 0.0 \\
      &                             & 
{\bf 1 $+$ 2 $+$ 3}
& SB & 81 &
-0.083& 0.135& 0.173 & 0.113& 2.5 & 19.4 &
0.012 &    0.137& 0.107& 0.070& 2.5 & 10.0 \\
\hline 
\end{tabular}
}